\documentclass[usenatbib]{mn2e}
\usepackage{graphicx}
\usepackage{url} 
\usepackage{wrapfig}
\usepackage{subfigure}
\usepackage{amsmath}
\usepackage{amssymb}
\usepackage{appendix}
\usepackage{alltt}
\usepackage{longtable}
\usepackage{tabularx}
\usepackage{ifpdf}
\ifpdf\else\usepackage{graphics}\fi

\makeatletter
\renewcommand{\fnum@figure}{{\bf Figure \thefigure}}
\renewcommand{\fnum@table}{{\bf Table \thetable}}
\makeatother
\let\ocaption\caption
\renewcommand{\caption}[2][]{\ocaption[#1]{{\small\it #2}}}

\title[RFI classification]{Post-correlation radio frequency interference classification methods}
\author[A.R. Offringa et al.]{
A.R.~Offringa$^1$,
A.G.~de~Bruyn$^{1,2}$,
M.~Biehl$^3$,
S.~Zaroubi$^1$,
G.~Bernardi$^1$,
\newauthor
V.N.~Pandey$^1$
\\
\small{$^1$Kapteyn Astronomical Institute, University of Groningen, PO Box 800, 9700 AV Groningen, The Netherlands} \\
\small{e-mail: \texttt{offringa@astro.rug.nl} } \\
\small{$^2$ASTRON, PO Box 2, 7990 AA Dwingeloo, The Netherlands} \\
\small{$^3$Institute for Mathematics and Computing Science, University of Groningen, P.O. Box 407, 9700 AK Groningen, The Netherlands}
}
\pagerange{\pageref{firstpage}--\pageref{lastpage}}
\date{Accepted 2010 February 3.  Received 2009 September 15}
\pubyear{2010}

\begin{document}
\label{firstpage}
\maketitle
\begin{abstract}
We describe and compare several post-correlation radio frequency interference classification methods. As data sizes of observations grow with new and improved telescopes, the need for completely automated, robust methods for radio frequency interference mitigation is pressing. We investigated several classification methods and find that, for the data sets we used, the most accurate among them is the \texttt{SumThreshold} method. This is a new method formed from a combination of existing techniques, including a new way of thresholding. This iterative method estimates the astronomical signal by carrying out a surface fit in the time-frequency plane. With a theoretical accuracy of 95\% recognition and an approximately 0.1\% false probability rate in simple simulated cases, the method is in practice as good as the human eye in finding RFI. In addition it is fast, robust, does not need a data model before it can be executed and works in almost all configurations with its default parameters. The method has been compared using simulated data with several other mitigation techniques, including one based upon the singular value decomposition of the time-frequency matrix, and has shown better results than the rest.
\end{abstract}


\section{Introduction}
Around 1980 when the radio spectrum was becoming more and more occupied as a result of technical advancements \citep{impact-of-warc79}, radio observers started to mitigate the radio frequency interference (RFI) caused by electronic equipment \citep*{interference-and-radioastronomy-1991}. Until recently, on-line thresholding and manual flagging of post correlated data used to be sufficient to suppress RFI artefacts in the data. However, as the volume of data and the required sensitivity of observations increased significantly, and the contamination of RFI through an increased usage of electronic equipment grew, new methods had to be developed to deal with the situation.

RFI mitigation can be applied in two different stages: a pre-correlation stage and a post-correlation stage. The pre-correlation mitigation stage is very powerful as the observational data is still available at its highest time resolution. For example, there are methods that blank or subtract short periodic radar RFI bursts on-line \citep*{pulse-blanking}, leaving the astronomical signal intact with only a very slightly increased signal to noise ratio. Any residual RFI has to be removed during the data reduction or imaging stage, which is often performed manually, for example by finding appropriate clipping levels for contaminated baselines until the reduced data is free of artefacts. Pre-correlation methods have to handle large amounts of data in a very short time and, because of hardware constraints, they can often only access limited dimensions of the data, such as the data from a single dish or station, or the data from a small time range.

Several methods from signal processing have been used to perform the first pre-correlation mitigation stage. Examples of these are thresholding using $\chi^2$-statistics \citep{chi-square-time-blanking-weber} or a Neyman-Pearson detector \citep{multichannel-rfi-mitigation}; spatial filtering with eigenvalue decomposition of a spatial correlation matrix \citep{multichannel-rfi-mitigation,hampson-spatial-nulling-2002} or by subspace tracking \citep{ellingson-spatial-nulling-2002}; the \texttt{CUSUM} method \citep*{wsrt-rfims}; and adaptive cancellation with a reference antenna \citep{adaptive-cancellation}. In the post-correlation phase, manual flagging is often the only option, but the use of an independent RFI reference signal to subtract the RFI \citep*{post-correlation-reference-signal}, fringe fitting \citep{fringe-fitting-rfi-mitigation} and post-correlation spatial filtering are possible. However, none of the above are applicable or sufficient in all cases or for all types of RFI.

In modern observatories that operate at low frequencies, such as the Westerbork Synthesis Radio Telescope (WSRT), the Giant Metrewave Radio Telescope (GMRT), the Low Frequency Array (LOFAR), and the Expanded Very Large Array (EVLA), RFI mitigation is an essential component in the signal processing. In the case of LOFAR, there are high sensitivity requirements, especially for the Epoch of Reionization project (\citet{lofar-foreground,lofar-simulations}), in what might be a busier RFI environment, with data sets up to a petabyte in size. RFI mitigation before correlation remains important \citep{lofar-rfi-requirements}, yet the amount of data will be too large for manual post-correlation flagging, implying the need for automated flagging strategies.

RFI comes in many forms \citep{rfi-mitigation-overview-fridman-baan,interference-model-lemmon}. The strong RFI that is problematic is often either local in frequency, such as RFI caused by television stations, aeroplanes and radar, or is local in time, e.g., broadband RFI caused by phenomena such as lightning, high-voltage power cables and electrical fences. Sometimes, the frequency of RFI drifts with time as shown later in Figure~\ref{fig:frequency-changing-rfi}. This can be caused by Doppler shifting of a satellite signal or by imperfect transmitters. A different class of RFI is caused by weakly transmitting, but stationary and therefore systematic, devices on site. This class of RFI is hard to recognize, as it might cover all the channels in a spectral band. In fringe stopping interferometers, the fringe rotation causes this type of RFI to have a sinusoidal response in the time-frequency domain \citep*{rfi-response-thompson-1982}. It can be recognized and subtracted in various ways, as for example described recently by \citet{fringe-fitting-rfi-mitigation}.

To select an RFI mitigation strategy, several considerations should be taken into account:
\begin{itemize}
 \item The true/false-positive ratio of the RFI classification;
 \item The speed of the algorithm;
 \item Detection or recovery, i.e., whether detection and flagging of contaminated areas is sufficient. In certain situations, it might be necessary to recover contaminated data, i.e., to subtract the RFI from the data;
 \item The effects of RFI mitigation on the noise. For example, a difference in the observed noise level caused by RFI will be fatal for the LOFAR Epoch of Reionization experiment \citep{lofar-foreground}.
\end{itemize}

In this paper we will evaluate the effectiveness of several automatic post-correlation RFI mitigation methods and their combinations. The methods will be compared with each other in order to be able to pick a general optimal post-correlation RFI strategy. We will do this by testing the methods on both artificial data and data from WSRT that has been observed in the frequency range of LOFAR. Testing the methods on WSRT data will also provide an indication of the effects of the RFI environment on future LOFAR observations.

In the next section we describe a new method of flagging RFI. We present our results, including the comparative study, in section \ref{results-chapter} and discuss the results in section \ref{conclusion-chapter}. In section \ref{future-work-chapter} we discuss some future directions for further work in this area.

\section{Methods} \label{methods-chapter}
Radio astronomers have developed their own ways of dealing with RFI during data reduction using numerous astronomical software packages. In many cases, this implies flagging by hand -- a tedious and time consuming job. Many toolkits, such as {\sc aips}\footnote{{\sc aips}: Astronomical Image Processing System, \\\url{http://aips.nrao.edu/}.}, {\sc aips++}\footnote{{\sc aips++}, sequal of {\sc aips}, \url{http://aips2.nrao.edu/}.}, {\sc miriad}\footnote{{\sc miriad}, a data reduction package tailored for the Australia Telescope Compact Array (ATCA), \\\url{http://www.atnf.csiro.au/computing/software/miriad/}.} and {\sc newstar}\footnote{{\sc newstar}, a data reduction package tailored for the Westerbork Synthesis Radio Telescope \citep{newstar}.}, provide specific features to perform flagging, such as the \texttt{FLAGR} task in {\sc aips++}. Astronomers have automated the process further by designing scripts in which common signal processing techniques such as thresholding, smoothing, line detection and curve fitting are combined \citep*{statistical-rfi-removal, median-filtering-bath-2005}. Another common signal processing technique known as Singular Value Decomposition has recently been used for the automatic removal of broadband RFI in GMRT observations \citep{the-gmrt-eor-experiment}. We will describe some of the techniques available that relate to a new method of interference mitigation that we will introduce, and finally we will explain the new method itself.

\subsection{Post-correlation thresholding}
Since RFI increases the measured absolute amplitude of a signal, thresholding is an effective method that is often used to remove strong RFI. The threshold level is often globally determined, or sometimes set relative to the variance or mode distribution parameters per baseline. These can be stably estimated using, for example, the Winsorized variance or mode \citep{variance-estimates}. All values that are outside a certain range around the mean or median are flagged as bad data and not used in subsequent data reduction. Sometimes a number of samples around a bad data sample are flagged as well. Most astronomical reduction toolkits provide options to threshold part of a data cube, allowing different thresholds at the cost of an increased effort for the astronomer. An important consequence of thresholding is that good data is selected with a bias. When many non-contaminated samples are above the threshold, they will be flagged and not used in subsequent data reduction. As a result, artefacts such as incorrect flux densities might be caused in the image plane. It is therefore important to have a low false-probability rate of RFI detection.

\subsection{Surface fitting and smoothing} \label{curve-fitting}
A surface fit to the correlated visibilities $V(\nu,t)$ as a function of frequency $\nu$ and time $t$ can produce a surface $\hat{V}(\nu,t)$ that represents the astronomical information in the signal. Requiring $\hat{V}(\nu,t)$ to be a smooth surface is a good assumption for most astronomical continuum sources, as their observed amplitudes tend not to change rapidly with time and frequency, whereas specific types of RFI can create sharp edges in the time-frequency domain. Because of the smoothing in both time and frequency direction, this method is not directly usable when observing strong line sources or strong pulsars. The residuals between the fit and the data contain the system noise $N_{\textrm{noise}}(\nu,t)$ and the RFI, $N_{\textrm{RFi}}(\nu,t)$, which can then be thresholded without the chance of flagging astronomical sources that have visibilities with high amplitude.

Several suitable surface fitting methods exist. As an example, in \citet{statistical-rfi-removal} a pipeline is described in which a two-dimensional, low order, dimensional independent polynomial is iteratively fitted to time-frequency tiles in the data using a least square fit:
\begin{equation}
 \hat{V}_{k}(\nu,t) = \sum_{i=1}^{N_\nu} a_{k,i} \nu^i + \sum_{i=1}^{N_t} b_{k,i} t^i + c_k,
\end{equation}
where $\hat{V}_{k}$ is the fitted surface that represents the astronomical information in the $k$-th tile,
$N_\nu, N_t$ are the polynomial order for the frequency and the time, respectively, and
$a_{k,i}, b_{k,i}, c_k$ are the coefficients of the fit for tile $k$.

The fit is performed iteratively, and values which have been flagged in previous iterations are excluded from the fit. This can be done by introducing a weight function $W_F(\nu,t)$, where $W_F(\nu,t)=0$ indicates that the value is flagged or outside the boundaries of the measured time or frequency range, and $W_F(\nu,t)=1$ means the value is accepted. The fit is performed by minimizing an error function $E_k$ for each tile: 

\begin{equation}
 E_k=\sum_{\nu} \sum_t W_F(\nu,t) f(\hat{V}_{k}(\nu,t), V(\nu,t))
\end{equation}
where $f(a,b)=a^2-b^2$ for a least squares fit or $f(a,b)=|a-b|$ for a fit with a minimal absolute error.

\begin{figure}
 \begin{center}
  \subfigure[$\hat{V}(\nu,t)$]{ \label{fig:tiled-fit}
   \includegraphics[width=25mm]{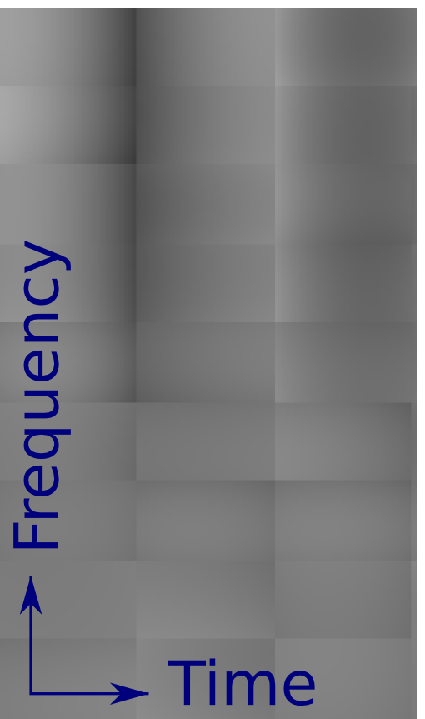}
  }
  \subfigure[$\hat{V}(\nu,t)-V(\nu,t)$]{ \label{fig:tiled-diff}
   \includegraphics[width=25mm]{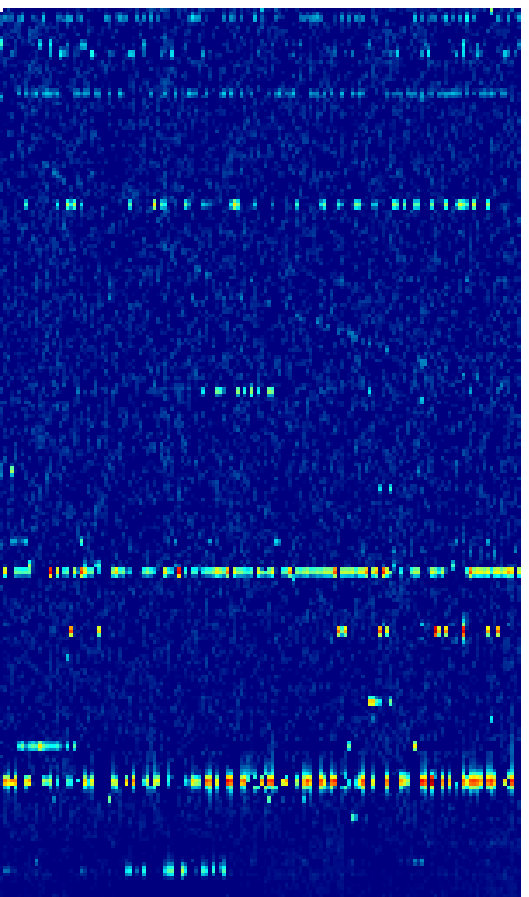}
   }
  \subfigure[Thresholded]{ \label{fig:tiled-thresholded}
   \includegraphics[width=25mm]{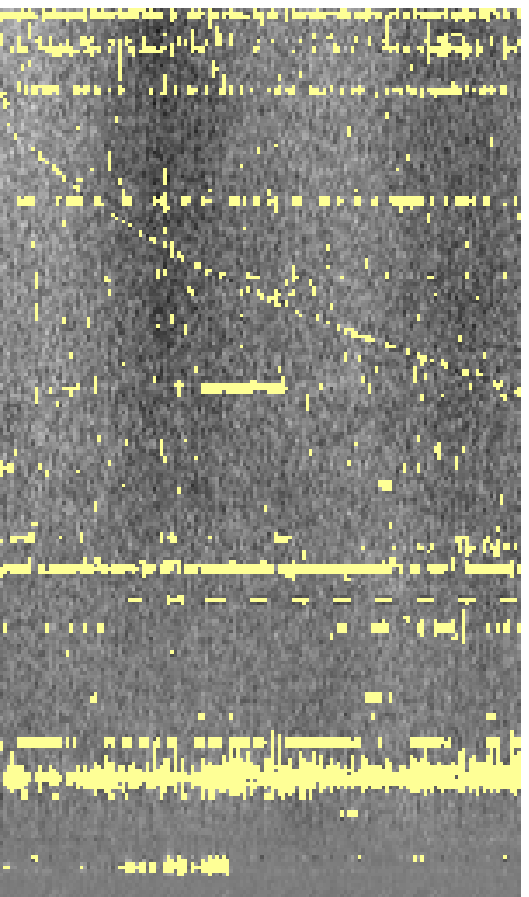}
   }
 \end{center}
 \caption{Tile-based polynomial fitting applied to the raw visibilities from an observation of 3C196 at 140 MHz using a 144m WSRT baseline (see \S\ref{wsrt-results}). Panel \subref{fig:tiled-fit} shows the tiled fit of the astronomical signal. Panel \subref{fig:tiled-diff} shows the difference between the fitted astronomical signal and the observed signal used for thresholding. Panel \subref{fig:tiled-thresholded} shows the flags on top of the original signal. The flags established by single pixel thresholding cover the RFI when verified by eye, although many false-positives can be seen which are caused by (``normal'') noise. The tile size used for this image is 30 frequency channels with 10 kHz width $\times$ 50 time scans with 10s integration time.}
 \label{fig:tiled}
\end{figure}

An example of this approach after a few iterations can be seen in Figure~\ref{fig:tiled}. In simple cases, the surfaces that are created with this approach represent the astronomical information reasonably well, and the method is also quite fast. However, as polynomial fits tend to show deviations near boundaries, the method is inaccurate near the boundaries of each tile.

Compared with tile-based approaches, sliding window methods tend to be more accurate. A simple example of a sliding window approach is to calculate the average of a window of size $N \times M$ around each data value:
\begin{small}
\begin{equation}
 \hat{V}(\nu,t) = \frac{1}{\textrm{\small{count}}} \sum_{i=-\frac{1}{2}N}^{\frac{1}{2}N} \sum_{j=-\frac{1}{2}M}^{\frac{1}{2}M} W_F \cdot V(\nu + i \Delta \nu,t + j \Delta t),
\end{equation}
\end{small}
with
\begin{equation}
 \textrm{count} = \sum_{i=-\frac{1}{2}N}^{\frac{1}{2}N} \sum_{j=-\frac{1}{2}M}^{\frac{1}{2}M} W_F(\nu + i \Delta \nu, t + j \Delta t)
\end{equation}

This method is still fast and creates a surface without tile edges. However, the sliding window average represents the astronomical signal less well. For example, peaks in the original function cause square-shaped edges in the fit, which in the end cause classification inaccuracies.

One way to overcome this problem is to calculate the local median instead of the local average. Values that have been flagged in a previous iteration should be ignored by the median calculation. The median is insensitive to peaks and the surface created by the local median remains smooth when the window is slid over the data. The median however is not always a good estimate of the sliding window centre sample specifically, as all samples have equal weight.

Another way to overcome the problem is to calculate a weighted average. Consider a weight function $W_d(i,j)$ that depends on the two components $i,j$ that represent the distances from the centre of the window in time and frequency respectively. Then
\begin{small}
\begin{eqnarray} \label{weighted-average} 
 \hat{V}(\nu,t) = \frac{ \sum_{i=-\frac{1}{2}N}^{\frac{1}{2}N}  \sum_{j=-\frac{1}{2}M}^{\frac{1}{2}M} 
 W_d(i, j) \left( W_F \odot V \right) \left(\nu_i,t_j\right) }{\textrm{\small{weight}}}
\end{eqnarray}
\end{small}
where
\begin{small}
\begin{equation} \label{weighted-total-weight}
 \textrm{weight} = \sum_{i=-\frac{1}{2}N}^{\frac{1}{2}N} \sum_{j=-\frac{1}{2}M}^{\frac{1}{2}M} W_d(i, j) W_F(\nu + i\Delta \nu, t + j \Delta t)
\end{equation}
\end{small}
This can be calculated very fast, since \eqref{weighted-average} is the convolution operation $W_d \ast \left( W_F \odot V \right)$ and \eqref{weighted-total-weight} is another convolution $W_d \ast W_F$, giving:
\begin{equation} \label{weighted-average-short}
 \hat{V} = \left(\left( W_F \odot V \right) \ast W_d \right) \oslash \left( W_F \ast W_d \right)
\end{equation}
where $\odot$ and $\oslash$ are the elementwise multiplication and division operators. A good choice for $W_d$ is the two-dimensional (dimensional independent) Gaussian function:
\begin{equation} \label{gaussian-weights}
W_d(i,j)=\exp \left(- \frac{i^2}{2\sigma_\nu^2} - \frac{j^2}{2\sigma_t^2} \right)
\end{equation}
Together, equations \eqref{weighted-average-short} and \eqref{gaussian-weights} essentially describe a weighted Gaussian smoothing operation, or more specifically, a Gaussian smoothing operation with missing data. The parameters $\sigma_\nu$ and $\sigma_t$ can be used to specify the level of smoothing in frequency and time respectively. Since the weight function is dimensionally separable, the convolutions can be dimensionally separated:
\begin{equation} \label{weighted-average-separated}
 \hat{V} = \frac{\left( W_F \odot V \right) \ast U_\nu \ast U_t }{ W_F \ast U_\nu \ast U_t}
\end{equation}
with $U_\nu(i) = W_d(i,0)$ and $U_t(j) = W_d(0,j)$. Each of the convolutions in \eqref{weighted-average-separated} is a one dimensional convolution, and this is therefore a fast operation.

\subsection{The cumulative sum method}
The cumulative sum (\texttt{CUSUM}) method is a well known analysis method used to detect changes in distribution parameters \citep{cusum,cusum-detection-of-abrubt-changes}, such as in quality control in production environments. If the cumulative sum of sequential samples exceeds an adaptive threshold, the system enters an alarmed state and changes can be made to correct the quality. In its common form, the likelihood for two distribution parameters is used to compute the threshold.

To turn this method into an RFI mitigation strategy, the total observed power or power received at a certain frequency by a single dish can be used as the sequential input values to the \texttt{CUSUM} method. The likelihoods of either variance or mean of RFI can be estimated using the variance of the signal \citep{cusum-cleaning-fridman-1996, wsrt-rfims}. Observing can be stopped as soon as RFI is detected, and can continue when reception has returned to normal. This method can be easily implemented for on-line RFI detection, as it is simple and fast. However, the \texttt{CUSUM} method does not estimate the start time of the change, it only detects the change quickly, which nevertheless may cost time and thus some bad data may leak through before the method detects faint RFI. Hence, the method is more applicable to a first check on the data than to actually perform flagging. The subsequent sections will describe a method that combines the detection strength of the \texttt{CUSUM} method with the possibility of performing flagging off-line.

\subsection{Combinatorial thresholding} \label{combinatorial-thresholding}
RFI bursts often affect multiple samples which are connected either in frequency or time. We will now introduce a new threshold mechanism that makes use of this knowledge: we will flag a combination of samples when a property of this combination exceeds some limit. Assume that $A$ and $B$ are neighbouring samples. In normal thresholding, we will look at each of the samples $A$ and $B$ individually and flag one of them if it exceeds some ``single sample'' threshold $\chi_1$. For combinatorial thresholding, a new flagging criterion is added: if $A$ and $B$ do not exceed the single sample threshold $\chi_1$ individually, they can still be flagged when $A$ and $B$ \emph{both} exceed a somewhat lower threshold $\chi_2$. If not, they can be combined with a third neighbour, $C$, and thresholded at $\chi_3$, etc. The more connected samples are combined, the lower the sample threshold.

Given a set of strictly decreasing thresholds, $\left\{ \chi_i \right\}_{i=1}^{N}$, a value will be classified as RFI if it belongs to a combination of $i$ values in either the time or frequency direction in which all absolute values are above the threshold $\chi_i$. To determine whether a single sample $R(\nu,t)$ should be flagged because of an RFI sequence in the frequency direction, the following rule is applied:
\begin{eqnarray} \notag
 \textrm{flag}\nu_M(\nu,t) = \exists i \in \{0\ldots M-1\}: \forall j \in \{0\ldots M-1\}: \\
 \label{eq-varthreshold}
 \left|R\left(\nu + \left(i-j\right)\Delta\nu, t\right)\right| > \chi_M
\end{eqnarray}
where $M$ is the number of samples in a combination. The flagging rules for the time direction are correspondingly determined. Finally, a sample is flagged if any of the two rules is satisfied.
We will call this method the \texttt{VarThreshold} method.

We will show a simple example to demonstrate the method. Consider the following values:
\begin{equation}
R=\left(\begin{matrix}
 1 & 2 & 1 & 4 \\
 4 & 1 & 1 & 4 \\
 2 & 2 & 1 & 4 \\
\end{matrix}\right)
\end{equation}
Each row represents a frequency channel and each column represents a time scan. Assume the high values in the fourth column were caused by broadband RFI. When using a normal threshold $\chi=3$, all samples with value 4 would be thresholded, including one false-positive. However, if we used combinatorial thresholding, with $\chi_1=5$ and $\chi_2=3$, we would threshold only the three broadband RFI samples.

The above text suggests an implementation of this method by a procedure which iterates over all samples and, for each sample, checks if it and its $M \in \mathcal{M}$ neighbours form an RFI sequence in one of the directions. Alternatively, an implementation can start by marking all samples above a certain $\chi_M$ as candidates. Subsequently, only the marked candidates that form a connected segment with more than $M$ connected samples in an orthogonal line in one of the directions are flagged. This procedure is repeated for all $M \in \mathcal{M}$. From this perspective, it is easy to add other morphological constraints. Instead of looking for straight lines in the time and frequency direction, an enhanced version might flag connected shapes covering a specific area, or shapes that form a line or curve in the plane, possibly not connected, that are likely to be caused by RFI.

\subsubsection{\texttt{VarThreshold} parameters}
The following list of parameters need to be optimised to make efficient use of this approach:
\begin{itemize}
 \item The false-positive rate on uncontaminated samples. The lower the value, the more RFI remains. The higher the value, the more uncontaminated samples will be flagged. We will discuss this in \S\ref{varthreshold-false-positive}.
 \item A set that defines which samples are combined. For this we define $\mathcal{M}$, a set containing the number of samples that will be combined in each of the four directions. Ideally, each sample will be combined with all samples of either the same frequency or the same time, i.e., $\mathcal{M} = \left\{ i \in \mathbb{Z}: 1 \le i \le \max(N_\nu, N_t) \right\}$, with $\mathbb{Z}$ the set of integers. Empirically, a small subset $\mathcal{M} = \left\{1,2,4,8,16,32,64\right\}$ works almost as well and saves summing and comparing many samples.
 \item The set of thresholds $\left\{\chi_M:M \in \mathcal{M} \right\}$ for the different number of combinations $M$. The total set of thresholds is expressed by two parameters, $\chi_1$ (the threshold on a single sample) and $\rho$, using the following formula:
\begin{equation} \label{eq-chiset}
  \chi_i = \frac{\chi_1}{\rho^ {\log_2 i }}
\end{equation}
A value of $\rho=1.5$ empirically seems to work well for the \texttt{VarThreshold} and the below defined \texttt{SumThreshold} method. To find $\chi_1$ for a desired false probability rate, $\rho$ is kept constant and the $\chi_1$ value is binary searched by performing mitigation on data selected from the distribution of the noise, with the values $\left\{ \chi_i \right\}_{i \in \mathcal{M}} $ computed as in \eqref{eq-chiset}, until the false probability rate is close to the desired rate.
\end{itemize}
Since the method is combined with a surface fitting strategy, the following parameters are added:
\begin{itemize}
 \item The number of iterations to be performed. The resulting accuracies are good with about 5 iterations.
 \item The iteration sensitivity as a function of the iteration number, $\eta(i)$. In each iteration, the threshold sensitivity is increased (more samples are flagged). To accomplish this, all the thresholds $\{\chi_i\}_{i \in \mathcal{M}}$ are decreased by dividing them by a factor of $\eta(i)$. Only during the last iteration will a sensitivity of 100\% be used. By slowly increasing the sensitivity a first bad fit to the background won't have much effect, since only the very strongly RFI contaminated samples are removed. Using an exponential function for $\eta(i)$ was found to work well. 
\end{itemize}

\subsubsection{The \texttt{VarThreshold} false-positive ratio} \label{varthreshold-false-positive}
Assume that $R \sim \mathcal{D}(\sigma_{N_s})$, where $R$ is the residual of the complex correlated visibilities $V$ and the surface fit $\hat{V}$, and $\mathcal{D}$ is a distribution with parameter $\sigma$. The probability that a non-RFI contaminated sample from the residual is larger than $\chi$ can be determined with:
\begin{small}
\begin{equation}
 \forall \nu \forall t : P\left(|R(\nu,t)| \ge \chi \right) = \int\limits_{-\infty}^{-\chi} \varphi_{\sigma}(x)dx + \int\limits_{\chi}^{\infty} \varphi_{\sigma}(x)dx,
\end{equation}
\end{small}
where $\varphi(x)$ is the probability density function of the distribution $\mathcal{D(\sigma_{N_s})}$. Note that the term $\int_{-\infty}^{-\chi} \varphi_{\sigma}(x)dx$ is only relevant when the distribution contains negative values -- unlike the Rayleigh distribution -- and the values are thresholded above $\chi$ as well as below $-\chi$.

The combined threshold false-positive rates can best be calculated numerically, since an analytical calculation is rather complex, even for $\mathcal{M}$ with a single combined threshold $\chi_M$. This analytical calculation will be demonstrated for $M=2$. First it is assumed that any two samples, $R(\nu_1,t_1)$ and $R(\nu_2,t_2)$, are independent when they are not RFI contaminated. This is the case if the fit represents the astronomical data and system noise is uncorrelated. With this assumption, the probability $P_\textrm{false}$ for a single non-contaminated sample $R_1$ with $M=2$ to be flagged in one of the four combinations with its neighbours $R_{2 \ldots 5}$ can be calculated with:
\begin{small}
\begin{eqnarray} \label{varthreshold-false-ratio}
\notag
 P_\textrm{false}(\nu,t) & \\ 
\notag
   = & P{\big (}\textrm{flag}\nu_{M=2}(\nu,t) \hspace{1mm} \vee \hspace{1mm} \textrm{flag}t_{M=2}(\nu,t) {\big )}\\
\notag
   = & P(|R_1| > \chi \wedge \exists i \in [2 \ldots 5]: |R_i| > \chi) \\
   = & P(|R| > \chi) - P(|R| > \chi) \left( 1 - P(|R| > \chi) \right)^4.
\end{eqnarray}
\end{small}
The corresponding formulae for larger M are more complex. When $\mathcal{M}$ contains more than one element, the false-positive ratios for the elements $M_i$ can not be simply added to obtain the combined false-positive ratio, as $P(\textrm{flag}\nu_{M_i}(\nu,t))$ and $P(\textrm{flag}\nu_{M_j}(\nu,t))$ are not statistically independent: both will at least make use of sample $R(\nu,t)$. Given this, the analytical expression becomes rather complex and the probability is evaluated numerically.

\begin{figure}
 \begin{center}
  \includegraphics[width=80mm]{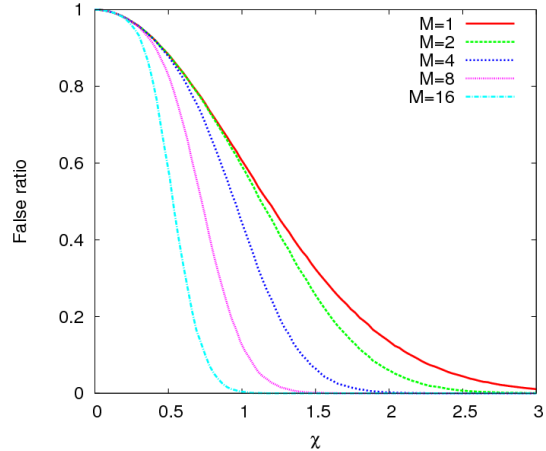}
 \end{center}
 \caption{The false-positives of the \texttt{VarThreshold} method when flagging with a single combination $\mathcal{M}=\{M\}$ without surface fitting. Samples were selected from a Rayleigh distribution, which is the distribution of the visibility amplitudes. $\chi$ is relative to the mode of the distribution. }
 \label{fig:varthreshd-false-pos-ratio}
\end{figure}

Figure~\ref{fig:varthreshd-false-pos-ratio} shows the result of calculating the total false-positive ratio numerically, for several values of $M$.

\subsection{The \texttt{SumThreshold} method}
Now we will present a variation on the \texttt{VarThreshold} method that improves the classification performance. This method, named the \texttt{SumThreshold} method, is a flagging method that combines samples as in the \texttt{VarThreshold} method. In this alternative case, the sum of a combination of one or more other samples is used as a threshold criterion. As in the \texttt{VarThreshold} method, the threshold $\chi_M$ is variable and depends on $M$, the number of samples that are summed.

Unlike the \texttt{VarThreshold} method, this approach allows the flagging of a sequence of samples when it contains samples with values below the sequence threshold value. However, without an additional rule, there are situations in which this method might flag too many samples. For example, consider the sequence $\left[0,0,5,6,0,0\right]$ that represents a strong RFI contamination in two samples. When the \texttt{SumThreshold} method without a second rule is applied with average threshold values $\chi_1=7$, $\chi_2=5$, $\chi_3=4$, \ldots, $\chi_6=1.8$, all six values would be thresholded, as their average exceeds $6\chi_6$. The following rule is therefore added: the values are thresholded in the increasing order $\chi_1$, $\chi_2$, \ldots, $\chi_M$. When a lower threshold has already classified samples as RFI contaminated, the samples will be left out of the sum and replaced by the average threshold level. In the example case, the values $5$ and $6$ will be classified as RFI by the second threshold, and therefore will be replaced by $\chi_6$ when combining all the six samples. The average of the sequence for the sixth threshold is therefore calculated as $\left(0 + 0 + \chi_6 + \chi_6 + 0 + 0\right) / 6 = \frac{2}{6}\chi_6$. As a consequence, only the samples with values $5$ and $6$ are flagged.

\subsubsection{The \texttt{SumThreshold} false-positive ratio}
We calculate the theoretical false-positive ratio for $M=2$ as for the \texttt{VarThreshold} method. The probability $P(T_{\chi,1,2})$ that the sum of two independent random samples exceeds a certain value $\chi$ is given by:
\begin{small}
\begin{eqnarray}
 \forall \nu_1 \nu_2 t_1 t_2 :
  P(T_{\chi,1,2}) & = & P\left[R(\nu_1,t_1)+R(\nu_2,t_2) \ge \chi \right] \notag \\
  & = & P\left(\mathcal{D}(2\sigma_{N_s}) \ge \chi \right) \notag \\
  & = & \int\limits_{\chi}^{\infty} \varphi_{2\sigma}(x)dx \label{pair-wise-probability}
\end{eqnarray}
\end{small}
When thresholding the average of a combination of two samples, each sample will occur four times in a hypothesis test, once with each of its neighbours. On uncontaminated samples, the probability of a false-positive for each of these tests is given by \eqref{pair-wise-probability}. The probability for a false-positive with the four tests applied on each sample becomes:
\begin{eqnarray*}
P(T_{\chi,1 \times 4}) & = & P(T_{\chi,1,2} \vee T_{\chi,1,3} \vee T_{\chi,1,4} \vee T_{\chi,1,5})
\end{eqnarray*}

\begin{figure}
 \begin{center}
   \includegraphics[width=80mm]{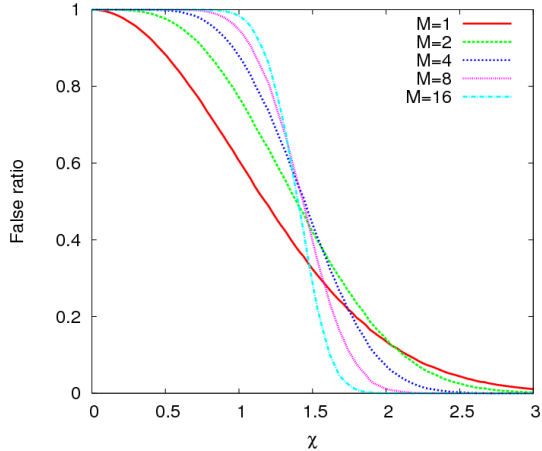}
 \end{center}
  \caption{The probability of a false-positive when thresholding with a single combination $\mathcal{M}=\{M\}$ using the \texttt{SumThreshold} method without surface fitting. The Rayleigh distribution was used for the simulation. $\chi$ is the average threshold relative to the distribution mode. Thus a combination of samples was thresholded when their sum exceeds $\chi \times M \times \sigma$. The false ratio for $M\ge2$ is different from the \texttt{VarThreshold} method (Figure~\ref{fig:varthreshd-false-pos-ratio}). Because of this difference, the parameter $\rho$ used to calculate the set of thresholds as in \eqref{eq-chiset} needs to be optimised for the methods individually. Although the false ratio is not decreased when comparing this method with the \texttt{VarThreshold} method, the true ratio is often increased (Figure~\ref{fig:roc-curves}).}
  \label{fig:sumthreshold-false-pos-ratio}
\end{figure}

Because the tests $\left\{T_{\chi,1,i}\right\}_{i=2}^{5}$ are dependent on each other, it is much easier to calculate the false-positive rates numerically. This can be performed by applying the \texttt{SumThreshold} on a large amount of data selected from the distribution $\mathcal{D}$. The result of such a simulation is in Figure~\ref{fig:sumthreshold-false-pos-ratio}.

\subsection{Singular Value Decomposition} \label{svd-method}
Singular value decomposition (SVD) is a mathematical tool for finding the singular values of a matrix, which can exhibit certain properties of the matrix.

A singular value decomposition consists of finding the complex unitary $M \times M$ and $N \times N$ dimensional matrices $U$ and $V$ containing respectively a left and right singular vector in each row, and the diagonal, $M \times N$ dimensional real matrix $\Sigma$ containing the singular values, such that:
\begin{equation}
 A = U \Sigma V^T
\end{equation}
RFI is mitigated from the data set by performing this decomposition on a matrix $A$. Each element $A_{ij}$ represents the measured flux, where $i$ is a baseline-frequency index and $j$ a time index. Each given matrix $A$ has a unique solution for the singular values $\Sigma$, if the singular values are sorted, but there is no unique solution for $U$ and $V$ (for example, $A$ remains equal when all values in $U$ and $V$ are negated). It is assumed that the highest singular values represent the singular values of the RFI data. To mitigate the RFI, the highest singular values in $\Sigma$ are set to zero and the new matrix $\hat{A}$ is recomposed from $U$, $\Sigma$ and $V$.

\begin{figure}
 \begin{center}
   \includegraphics[width=80mm]{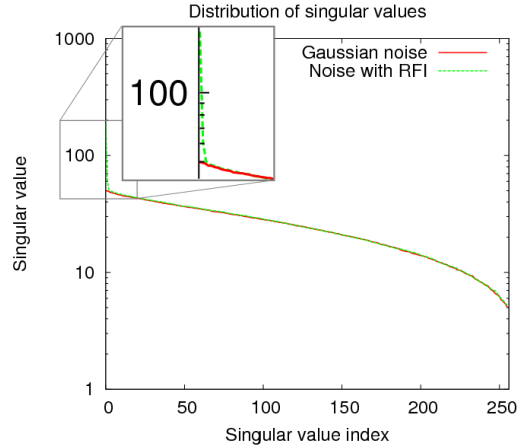}
 \end{center}
  \caption{The distribution of the singular values of two artificial measurements: one containing Gaussian noise only, the other containing Gaussian noise poluted by broadband RFI. In this example, the first five singular values are affected by the broadband RFI. In general, the number of singular values that are affected by RFI and the possibility to recognize them varies depending on the orthogonality properties of the RFI.}
  \label{fig:singularvalues}
\end{figure}

The number of singular values to be removed or set to zero has to be chosen in such a way that only the RFI is removed. The singular values that correspond to RFI are often strong outliers, whereas the singular values of Gaussian noise form a smooth curve. The position of the abrupt change in the curve of the singular values is used as the number of singular values to be removed, as is shown in Figure~\ref{fig:singularvalues}.

\subsubsection{Properties}
Let $L=\min(M,N)$, then:
\begin{equation}
 A_{ij} = \sum_{k=1}^L U_{ik} \Sigma_{kk} V_{jk}.
\end{equation}
$U$ and $V$ are unitary, $U\overline{U}=I$ with $\overline{U}$ the Hermitian transpose, and the rows and columns of the matrices form by definition a complex orthonormal basis. This implies:
\begin{equation} \label{svd-orthogonality}
 \forall i \in [1..M]:\sum_{j=1}^L U^2_{ij}=1.
\end{equation}
Hence there is at least one non-zero value in each row and column of the matrices $U$ and $V$, and setting a non-zero singular value to zero changes $A$. If $A$ contains real values only, equation~\eqref{svd-orthogonality} implies that all values in $U$ and $V$ are between $-1$ and $1$, and removing a singular value $\Sigma_{ii}$ can alter each value in $A$ by at most $\Sigma_{ii}$. In the complex case, removing a singular value can alter the absolute value of each value in $A$ at most by $\Sigma_{ii}$. In general, setting $\Sigma_{ii}$ to zero subtracts a matrix $\Gamma_i$ with rank 1 from $A$, as $\left( \Gamma_i\right)_{jk}=U_{ji} \Sigma_{ii} V_{ki}$, and thus all columns are linearly dependent.

The orthogonality properties imply that the order of the rows and columns in the original matrix $A$ do not change the singular values: the order of the rows and columns is irrelevant for the SVD method to detect RFI. Intuitively, the SVD method does not ``distinguish'' between a smoothly increasing amplitude, caused by astronomical sources, and RFI, and might fail to correctly subtract or detect RFI because of the astronomical signal.

If RFI is to be separated from the signal, the RFI and the signal have to adhere to the following properties:
\begin{itemize}
 \item All columns containing RFI (and consequently all rows) have to be orthogonal to the astronomical signal. In other words, for any column or row $\bmath{a}$ in the matrix, $\bmath{a}_\textrm{\small{RFI}} \cdot \bmath{a}_\textrm{\small{signal}} = 0$, with $\bmath{a}_\textrm{\small{RFI}}$ the RFI component and $\bmath{a}_\textrm{\small{signal}}$ the signal component in the data.
 \item The singular values of the RFI are substantially higher than the singular values of the astronomical signal. This requires the RFI to be strong.
 \item The individual RFI columns are either fully linearly dependent on or fully orthogonal to each other. If the individual RFI components are partially dependent, the largest part of the RFI is removed and the singular value of what is left of the RFI might not have enough 'gain' to be removed or flagged.
\end{itemize}

Iteratively fitting a surface and subtracting the surface, as in \S\ref{curve-fitting}, might improve the compliance to the first requirement, although it increases the execution time of the method. Another way to improve compliance to the requirement is to remove the astronomical signal by subtracting a good model beforehand.

It is useful to note that unitary transformations do not change the singular values of a matrix, although they might change the singular vectors. Since the Fourier transform is a unitary transformation according to Parseval's theorem, the following equation holds:
\begin{equation}
A=USV \Leftrightarrow \mathcal{F}(A)=U'SV'
\end{equation}
The consequence of this is that it does not matter whether the SVD method is executed in the time-frequency domain, the time-lag domain, or another Fourier domain, since setting singular values to zero in the Fourier domain would set the singular value to zero in the original domain.

\subsection{Input data types}
The combined thresholding methods described in this paper can be applied to several types of data: auto-correlated or cross-correlated, to specific polarizations or Stokes parameters, to amplitude or to phase, etc. 

We have compared flagging on cross-correlations and auto-correlations. The cross-correlations of each baseline can be processed with one of the flagging methods, resulting in $N(N-1)/2$ correlations to be processed. Alternatively, every antenna can be individually flagged by processing the auto-correlations, and samples in a baseline might be flagged if either of the corresponding samples in the individual antenna auto-correlations have been flagged. Only $N$ correlations need to be searched for RFI in this case. In addition to the benefit of speed, RFI is strongest in auto-correlations and the data contain no fringes from astronomical sources, as auto correlations do not have interference patterns, thus offering an improved accuracy in RFI detection. On the down side, some RFI might be present in auto-correlations that would have been mitigated by cross-correlation, and detecting RFI in auto-correlations potentially throws away some usable data in the cross-correlations.

In cases where the polarization of the observed electromagnetic waves is measured, the polarization might contain valuable information for RFI classification. For now, we have processed each polarization individually, without exploiting relationships between polarizations.

\section{Results} \label{results-chapter}

\subsection{Surface fitting results}

\begin{figure*}
 \begin{center}
  \subfigure[Original observation]{ \label{fig:frequency-changing-rfi}
   \includegraphics[width=80mm]{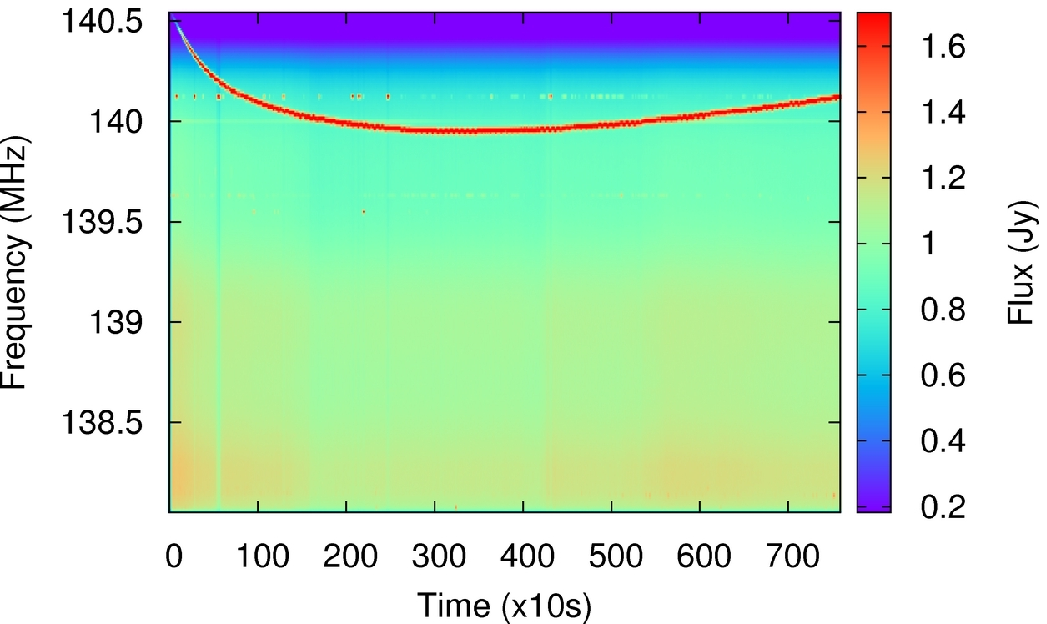}
  }
  \subfigure[After removing the highest singular values from the image (note the different flux scale).]{
   \includegraphics[width=80mm]{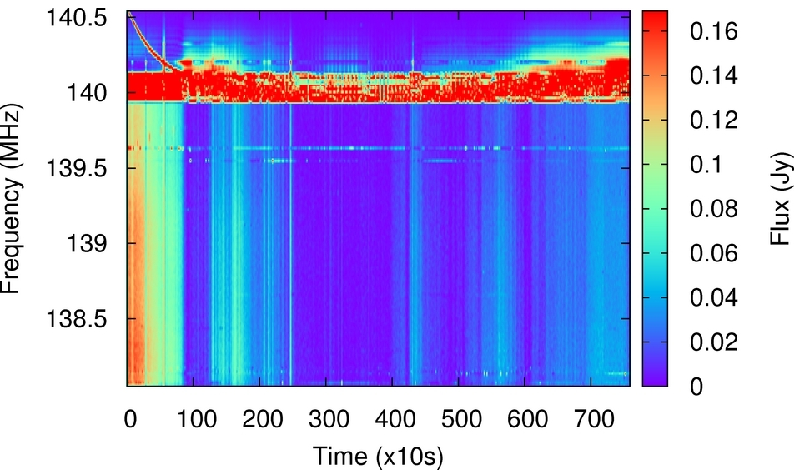}
   }
 \end{center}
 \caption{The auto-correlated data in this image demonstrate the inability of the SVD method to remove sources that slowly change frequency over time (e.g., because the source has a changing velocity in the direction of the antenna). This type of RFI seems to be relatively common in low frequency WSRT data. The RFI in this particular example is so strong that it can be easily removed by thresholding, but this plot is to illustrate the effects of such RFI. When the frequency-changing signal is faint and cannot be removed by thresholding, applying SVD will, as in this example, change the astronomical information in the data in an unpredictable way.}
 \label{fig:svd-diagonal-line}
\end{figure*}

\begin{figure*}
 \begin{center}
  \subfigure[Recomposed image from the low singular values.]{
   \label{fig:testsetH-svd-low}
   \includegraphics[width=112mm,height=25mm]{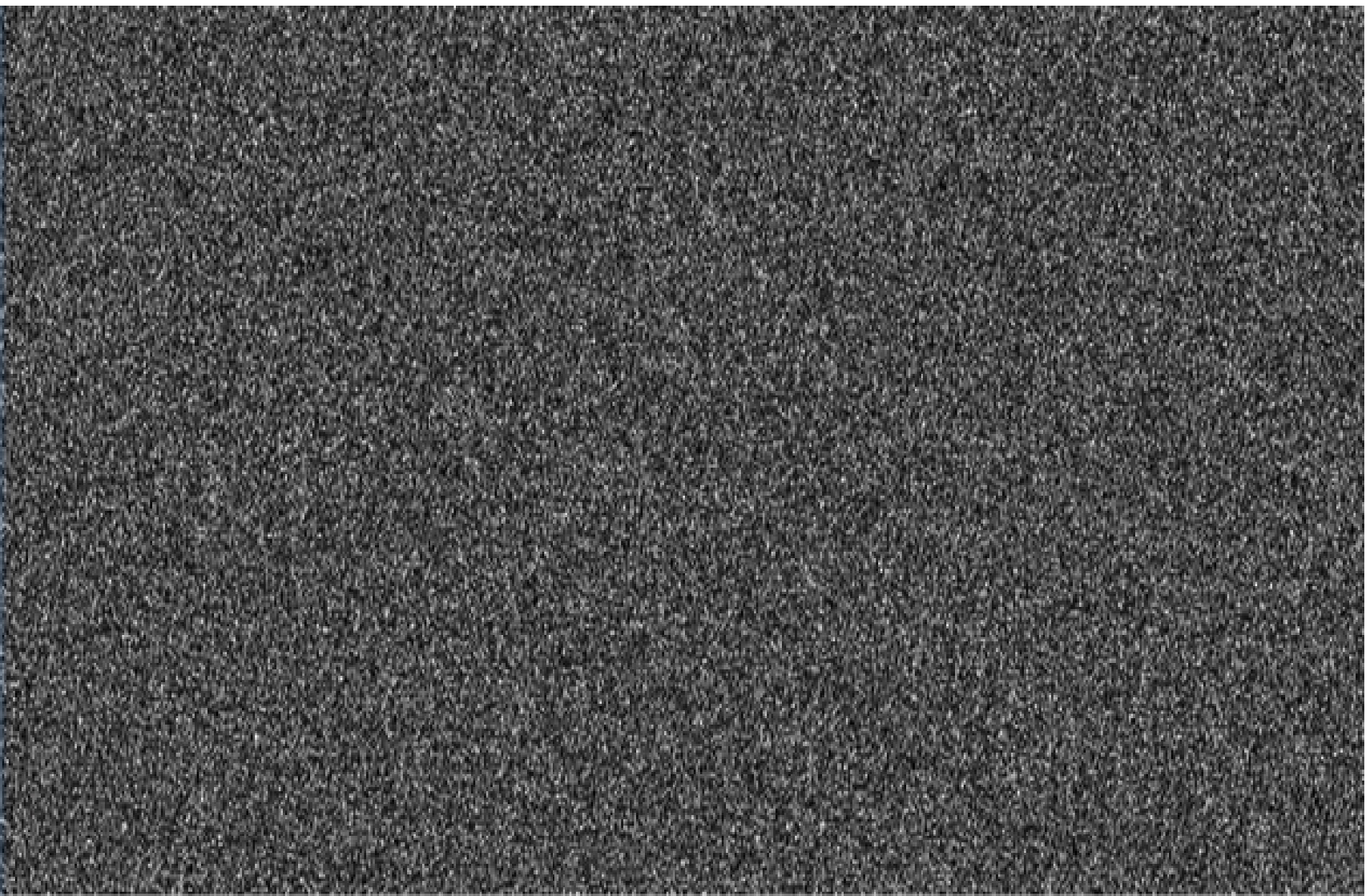}
   }
  \subfigure[Recomposed image from the high singular values.]{
   \label{fig:testsetH-svd-high}
   \includegraphics[width=112mm,height=25mm]{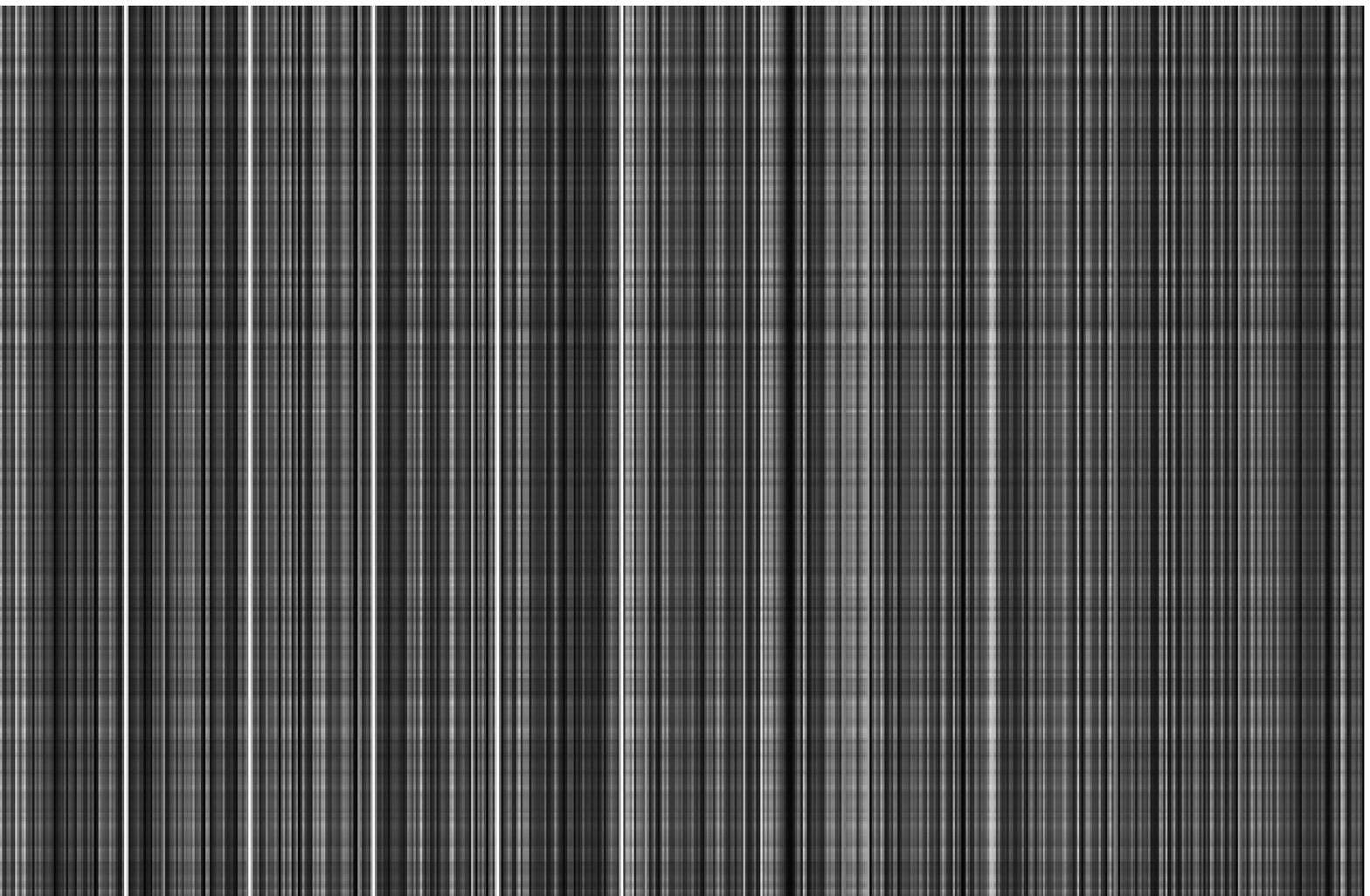}
   }
 \end{center}
 \caption{SV decomposition of test set A (Figure~\ref{fig:testsets-set-A}): noise with broad band RFI covering all channels homogeneously. The recomposed image from the low singular values (top panel) looks very promising: none of the RFI is left and the noise \emph{seems} to be untouched. However, a recomposition of the matrix with only high singular values (bottom panel), i.e., the part that has been subtracted from the image, shows that the noise is affected in an unpredictable way by the decomposition. This is the best case for the SVD method; in more realistic scenario's, the data should include a residual astronomical signal and broadband RFI that might not be linearly dependent.}
 \label{fig:testsetH-svd}
\end{figure*}

In \S\ref{curve-fitting} we described several surface fitting methods to estimate the astronomical signal in the frequency-time domain. We found that the surface fitting methods when combined with one of the classification methods do not differ much in accuracy. A sliding window approach was found to be more stable compared with a tile based approach. The Gaussian weighted average, a polynomial fit and the window median for the subtracted surface were found to be approximately equal in their accuracies after optimising their parameters such as the window size, the Gaussian kernel size and the order of the polynomial, although their parameters do influence the accuracy.

Finding global parameters that always work well (or automatic procedures to find the parameters) is not trivial. The algorithm can handle data with very different characteristics: it can be applied to XX, XY, YX or YY polarizations, auto- or cross-correlations from either long or short baselines, for LOFAR or for WSRT data, before or after calibration, etc. To use the same surface fitting parameters in all these different situations, the window size, and if applicable the Gaussian kernel size, needs to be rather small. The expected amplitude changes of celestial signals are usually much less in the frequency direction, and setting the window size larger in the frequency direction improves stability. We used a typical size of the sliding window of 40 frequency channels $\times$ 20 time scans and Gaussian kernel parameters of $\sigma_\nu=15$ and $\sigma_t=7.5$. The numbers are based on trials using different observed and artificial data sets. The parameters are relative to the number of channels and number of time steps. For WSRT data, a channel is 10 kHz wide and a time scan is 10 seconds long. LOFAR will have a 1 kHz $\times$ 1 second correlation output resolution. For best results, the length and width of the window should be about three times the Gaussian kernel size or larger.

\subsection{RFI classification results}
\begin{figure*}
 \begin{center}
  \subfigure[Test set A: noise with broadband RFI contaminating all channels, ordered from strong (left) to weak (right).]{
   \includegraphics[width=70mm]{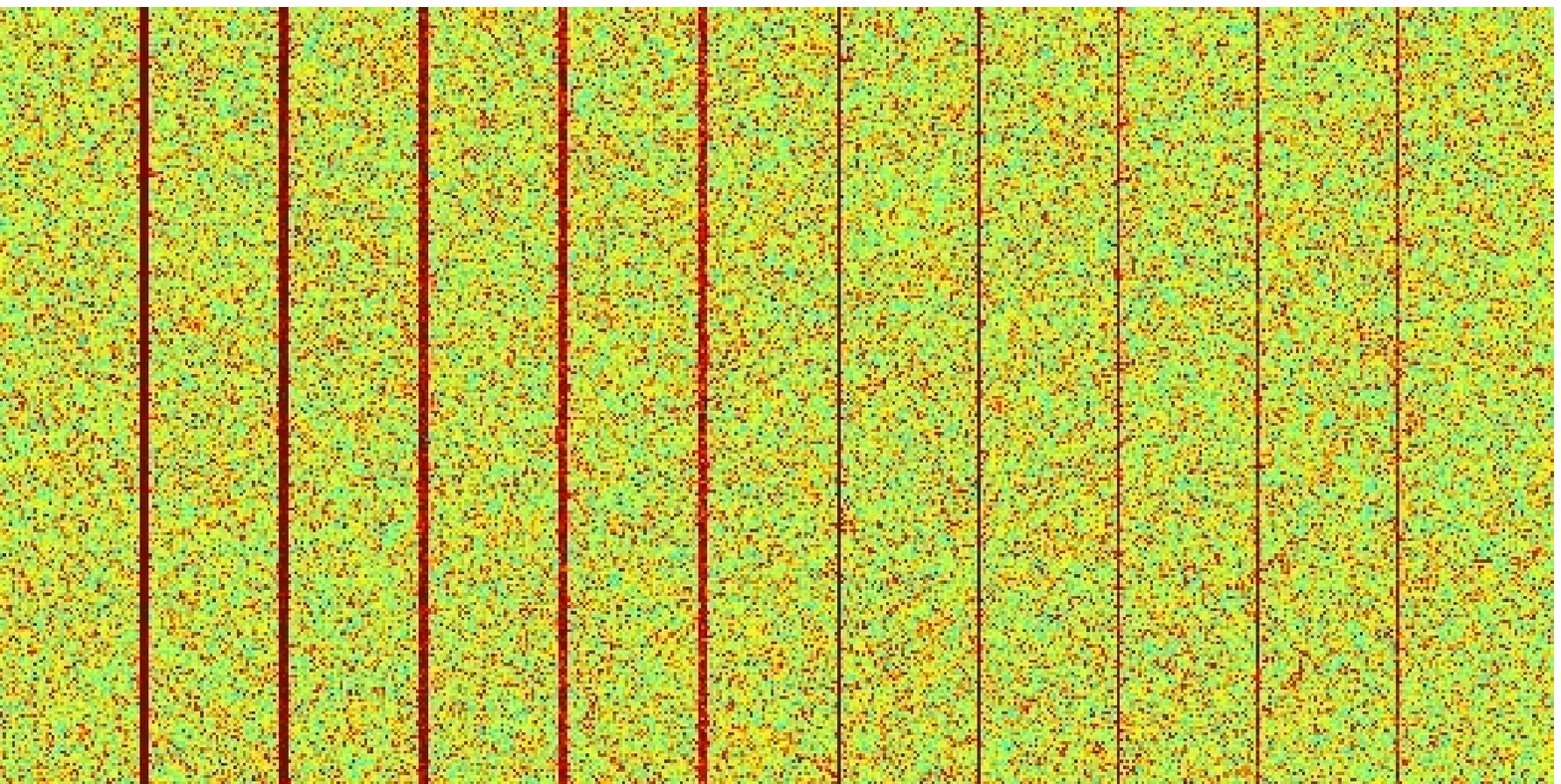}
   \label{fig:testsets-set-A}
   }
  \subfigure[Test set B: broadband RFI contaminating a part of the channels]{
   \includegraphics[width=70mm]{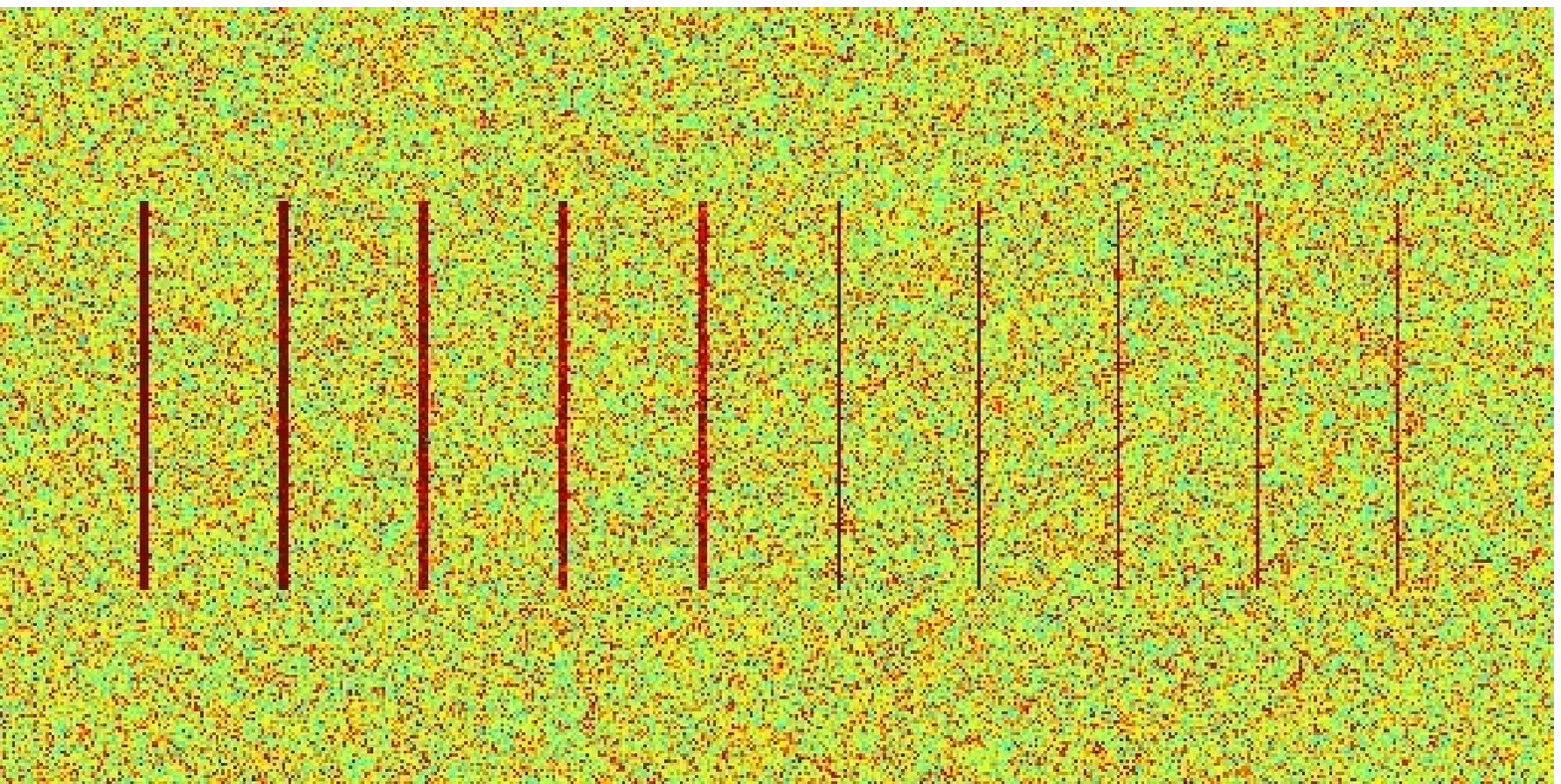}
   }
  \subfigure[Test set C: broadband RFI contaminating different channels]{
   \includegraphics[width=70mm]{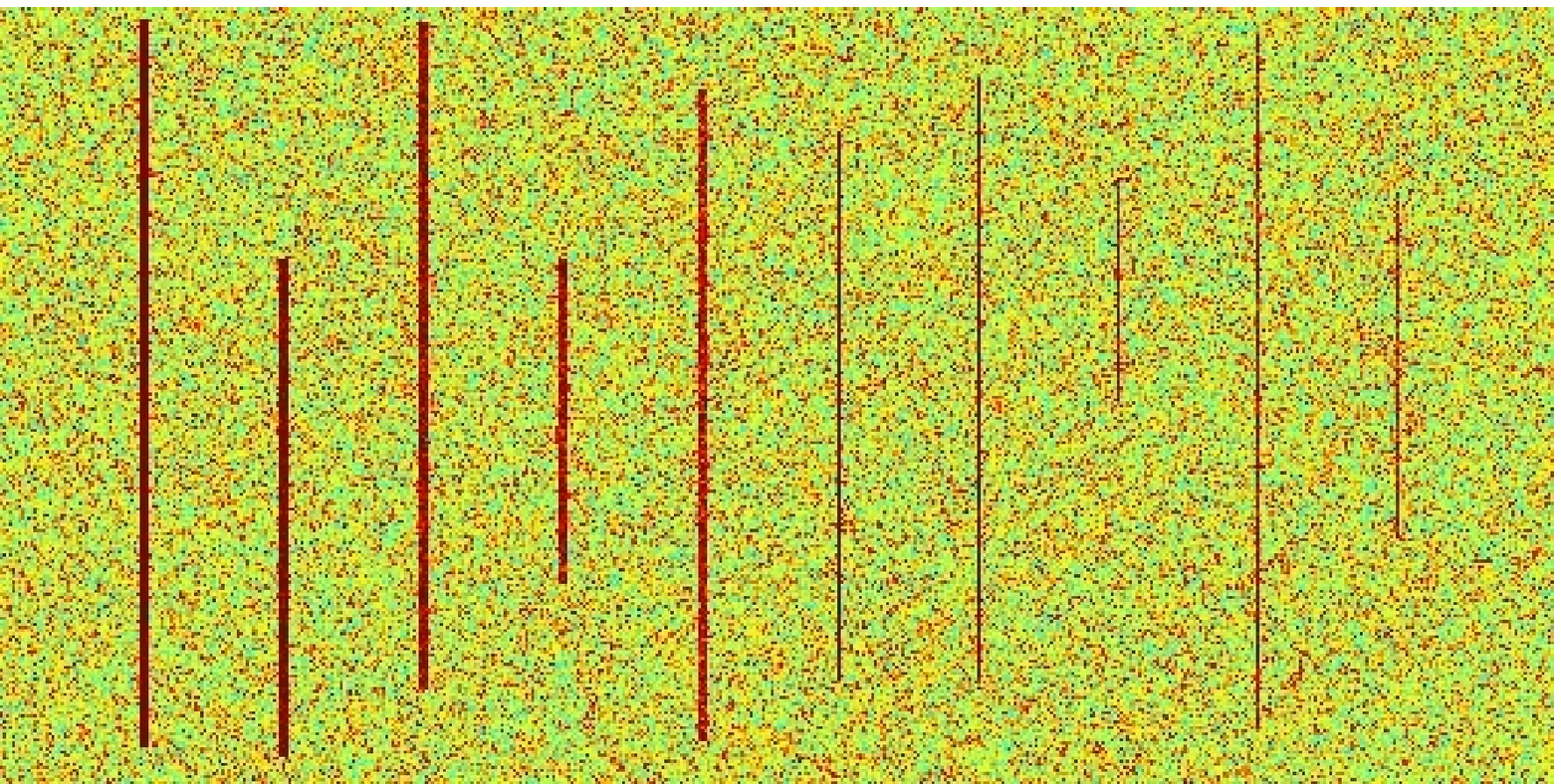}
   }
  \subfigure[Test set D: a simulated observation of the cross-correlation of three point sources being close together added to test set C]{
   \includegraphics[width=70mm]{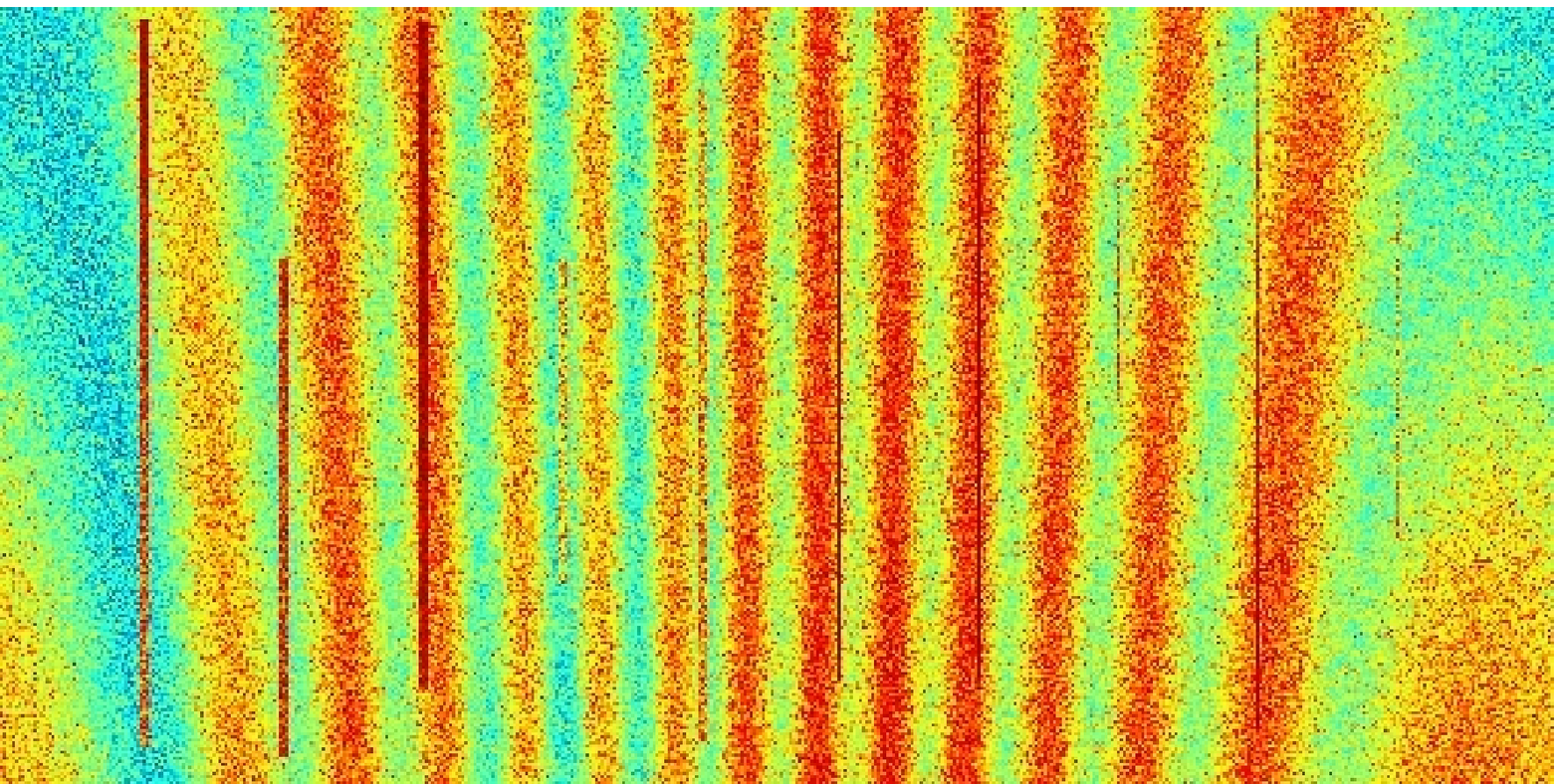}
   }
  \subfigure[Test set E: a simulated observation of the cross-correlation of five distant sources added to test set A]{
   \includegraphics[width=70mm]{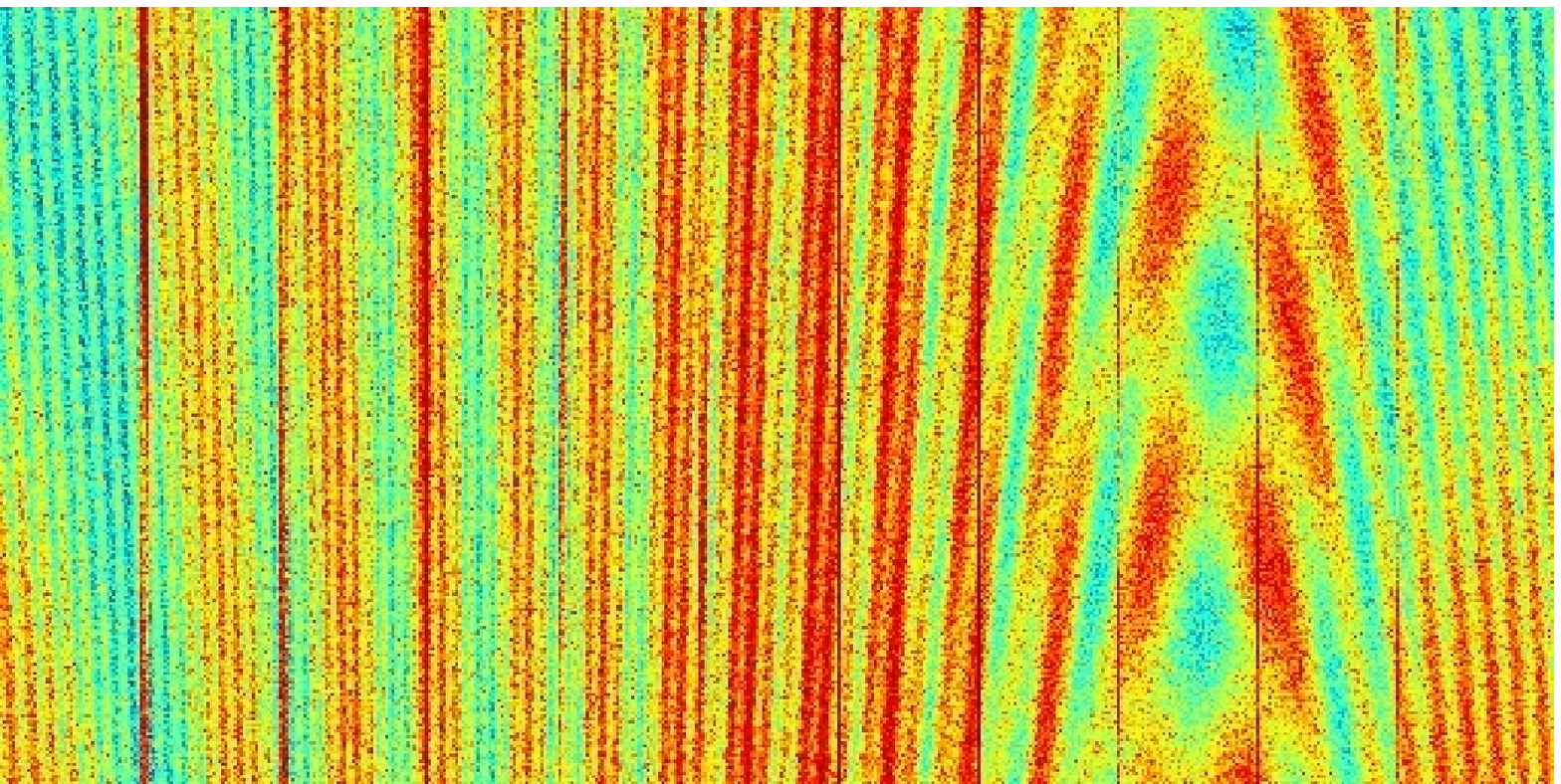}
   }
  \subfigure[Test set F: as E, but RFI as in test set C]{
   \includegraphics[width=70mm]{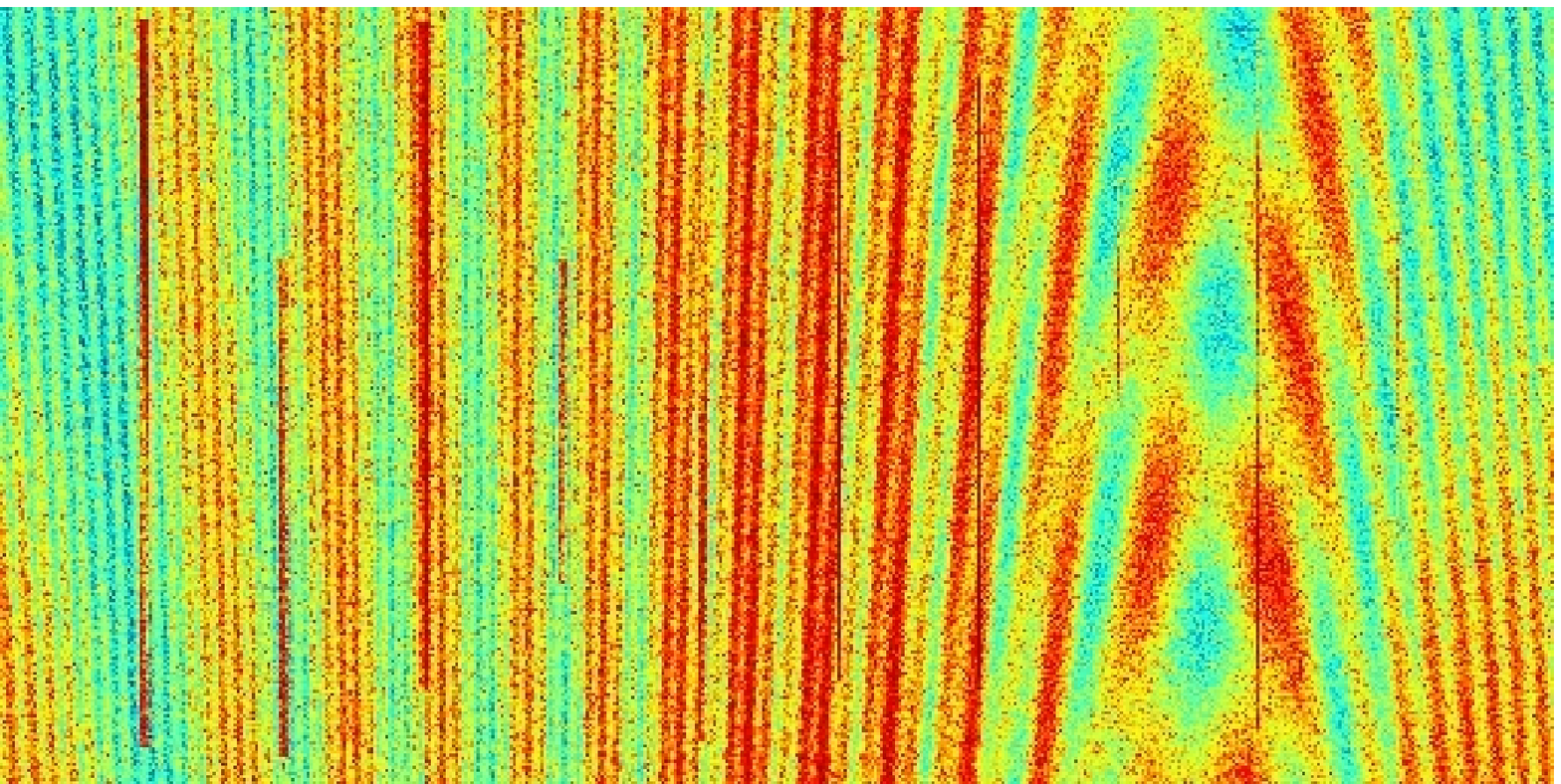}
   }
  \subfigure[Test set G: as F, but Gaussian smoothed before adding RFI]{
   \includegraphics[width=70mm]{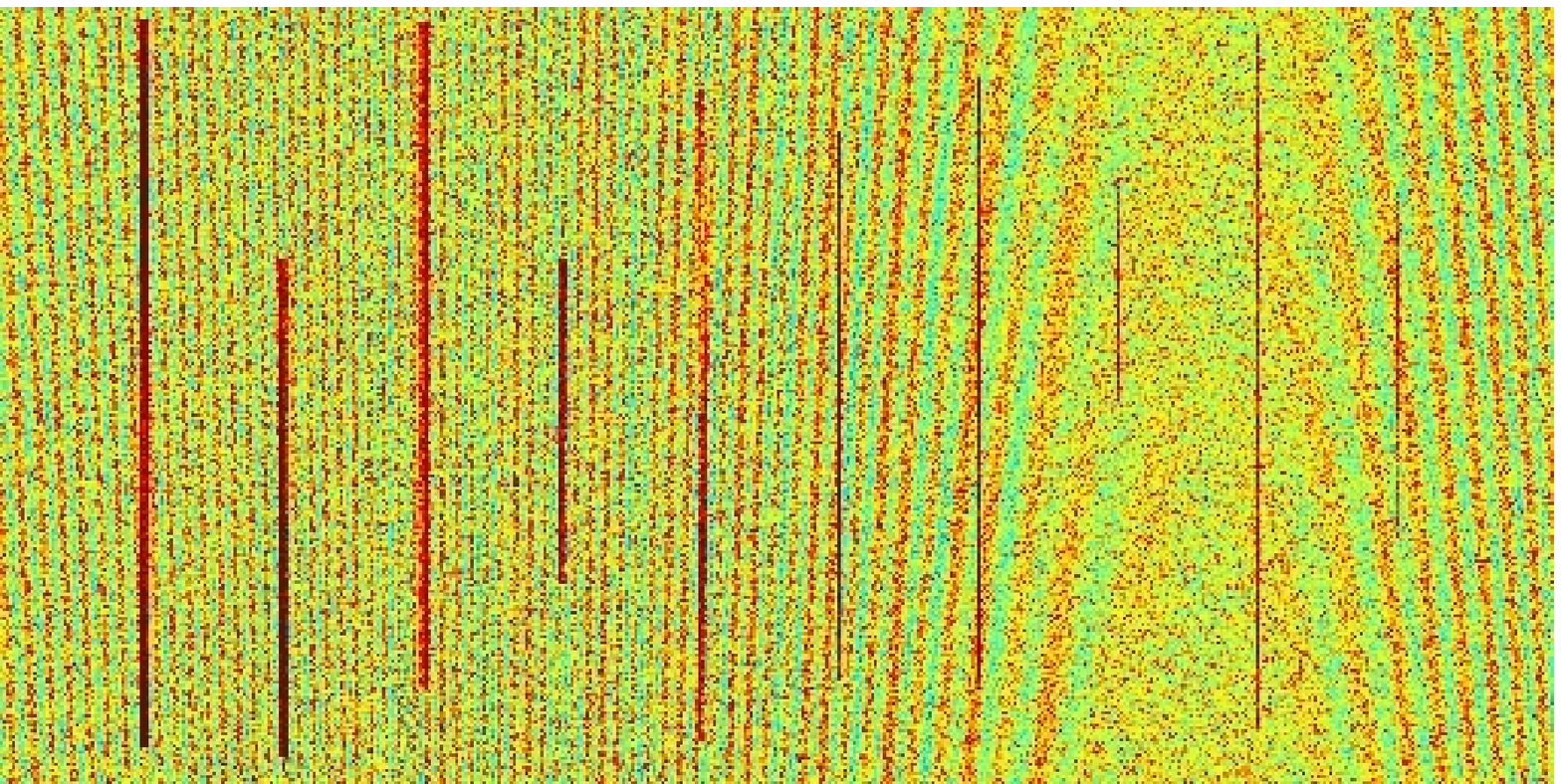}
   \label{fig:testsets-set-G}
   }
  \subfigure[Test set H: a high frequency background signal added to test set C]{
   \includegraphics[width=70mm]{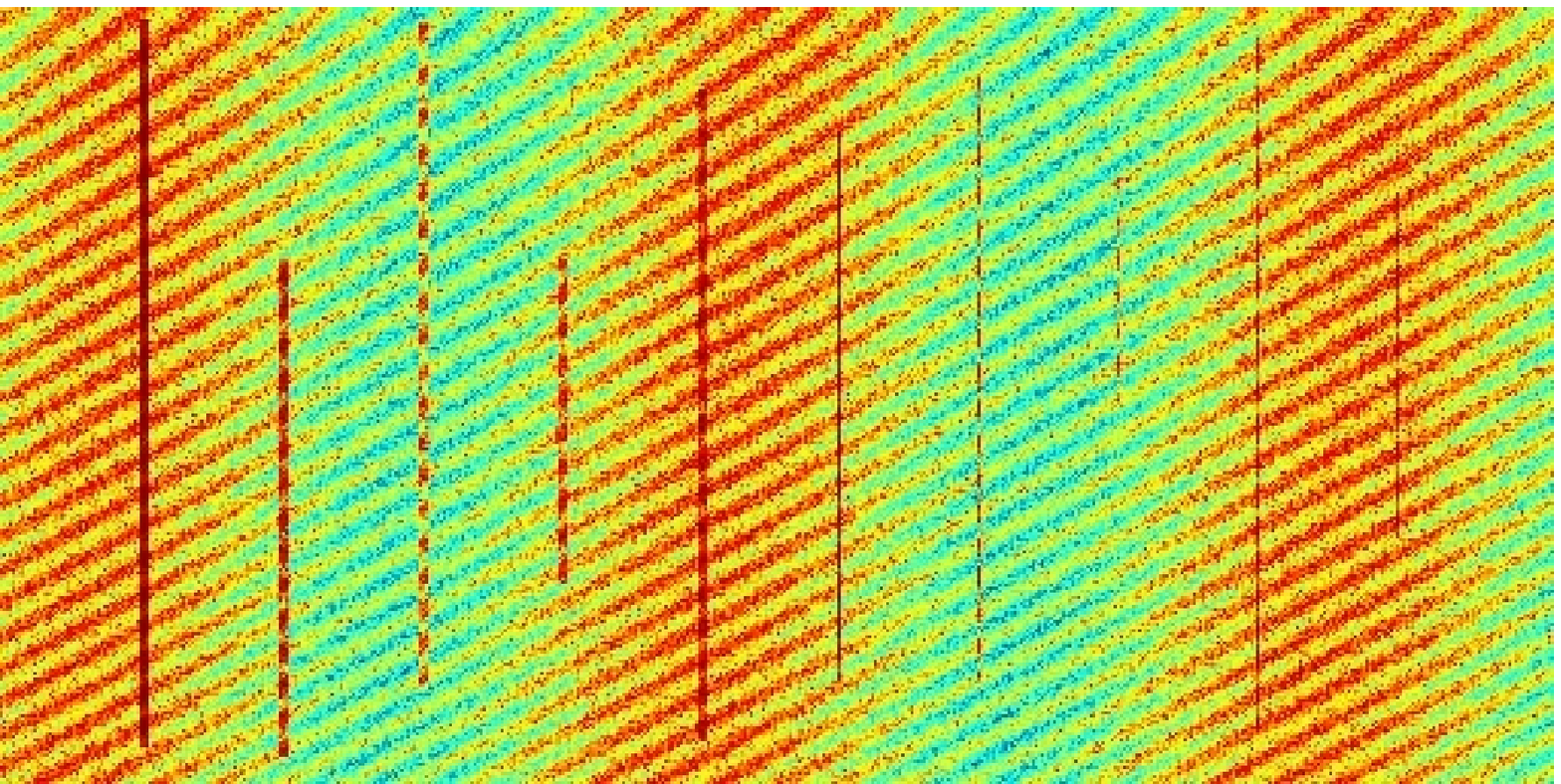}
   \label{fig:testsets-set-H}
   }
 \end{center}
 \caption{The artificial test sets containing broad band RFI, used for testing and parameter optimisation. In all images, time is along the horizontal and frequency along the vertical axis. All test sets simulate a similar baseline. }
 \label{fig:testsets}
\end{figure*}

\begin{figure*}
 \begin{center}
  \subfigure{
   \includegraphics[width=70mm]{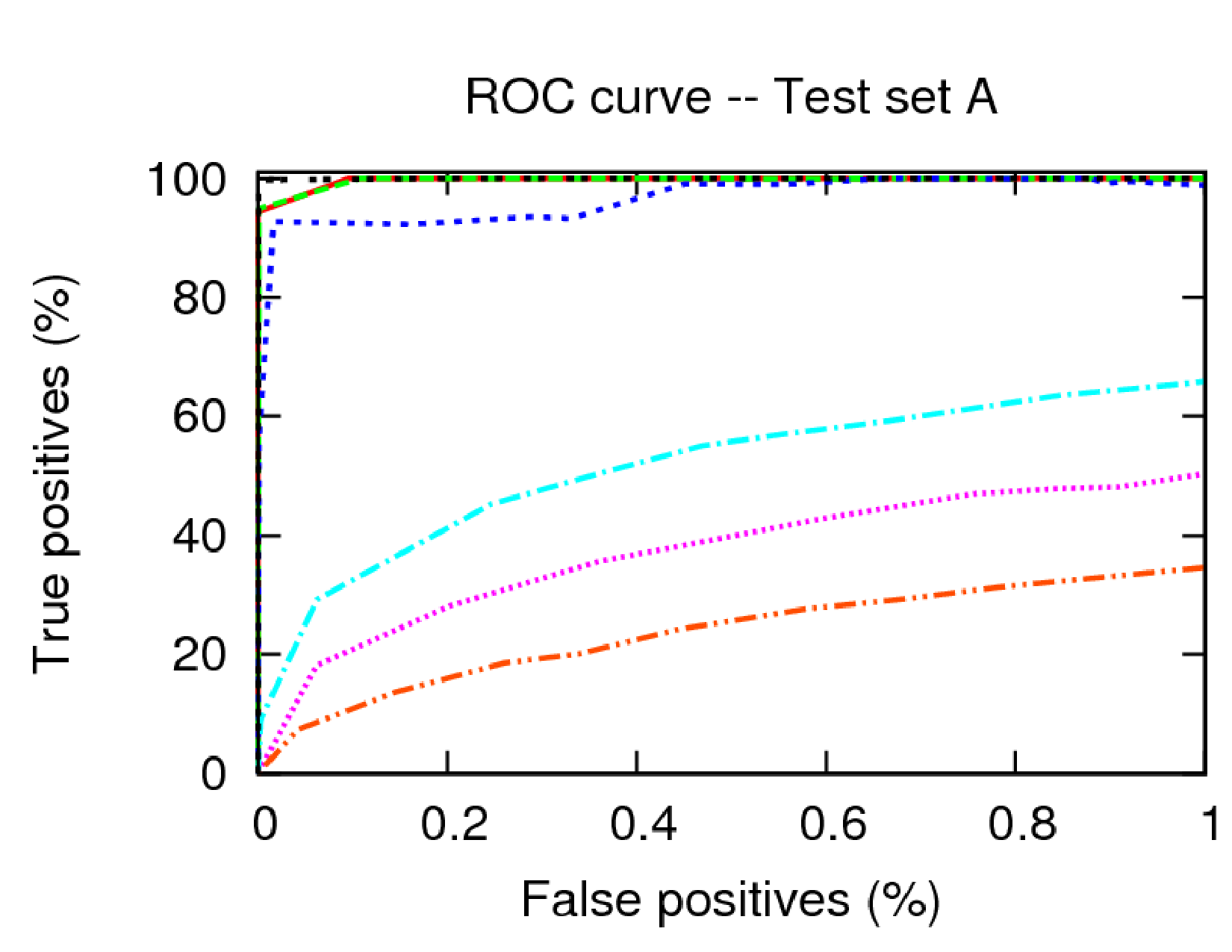}
  }
  \subfigure{
   \includegraphics[width=70mm]{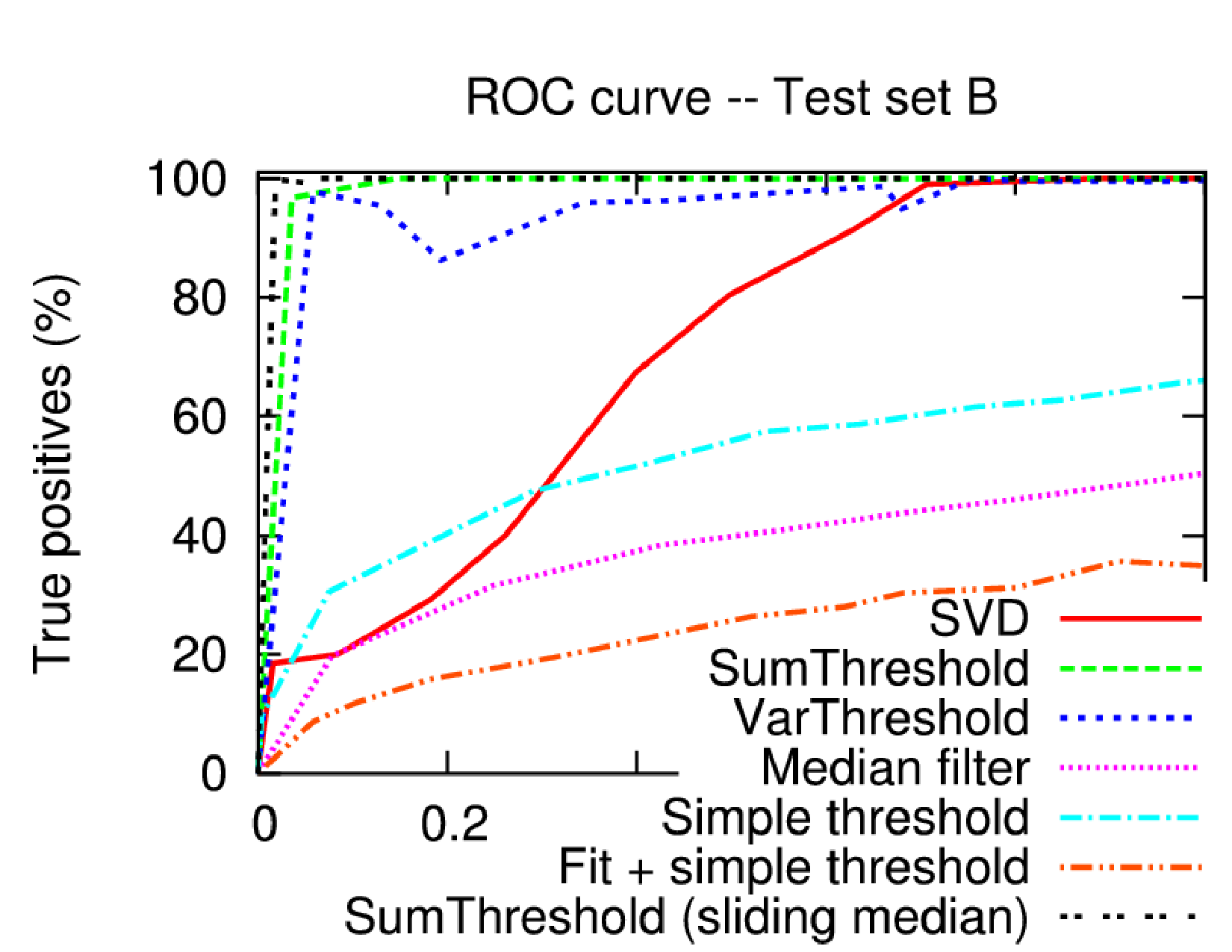}
  }
  \subfigure{
   \includegraphics[width=70mm]{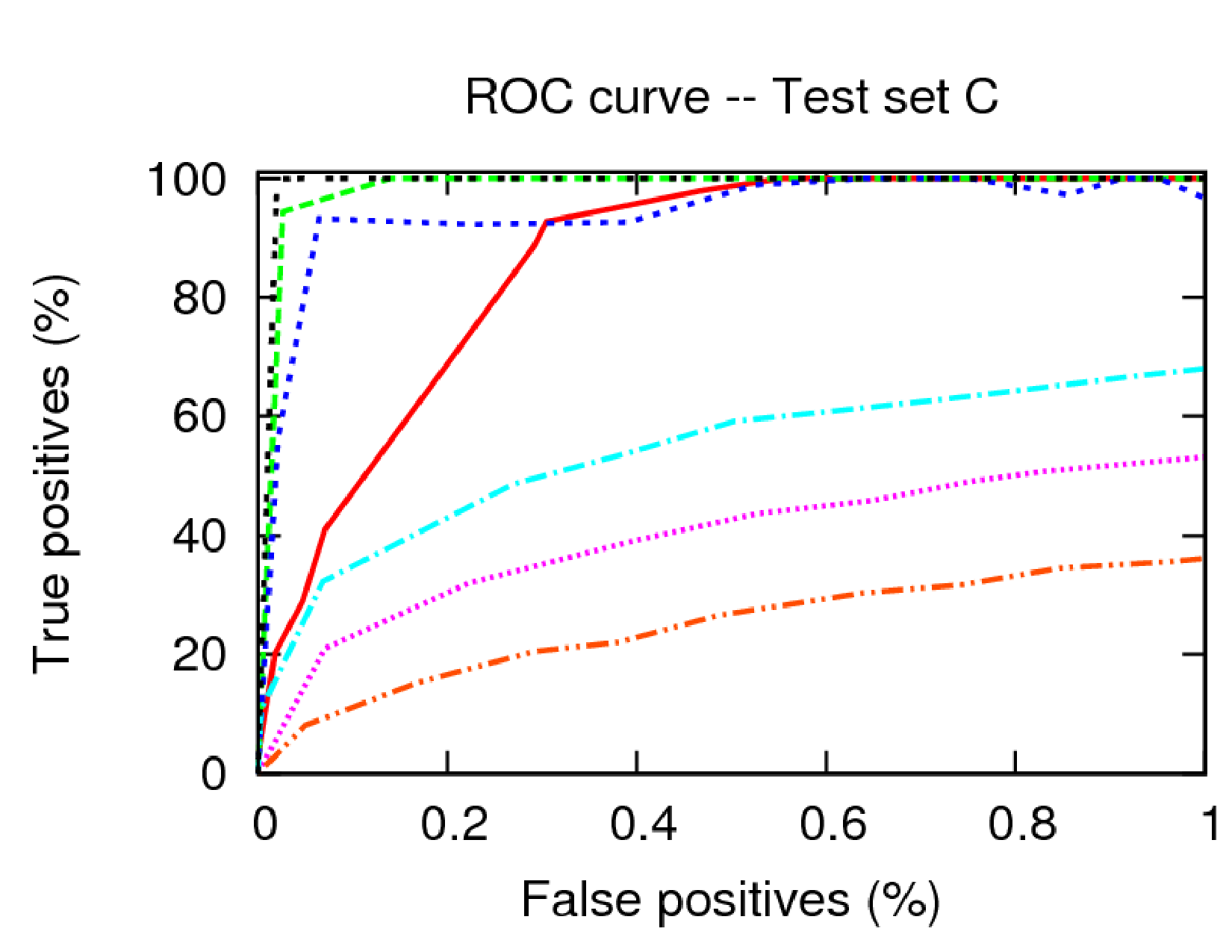}
  }
  \subfigure{
   \includegraphics[width=70mm]{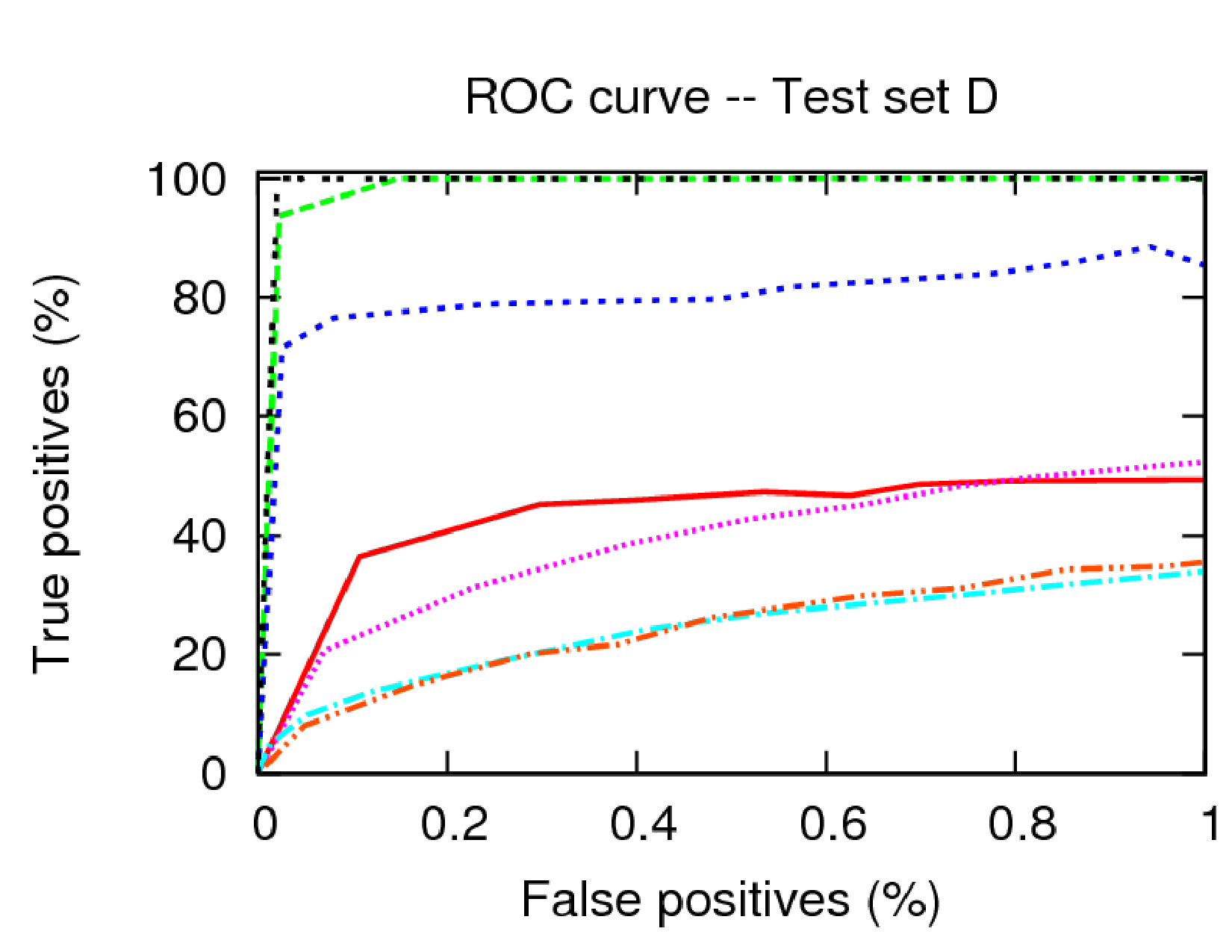}
  }
  \subfigure{
   \includegraphics[width=70mm]{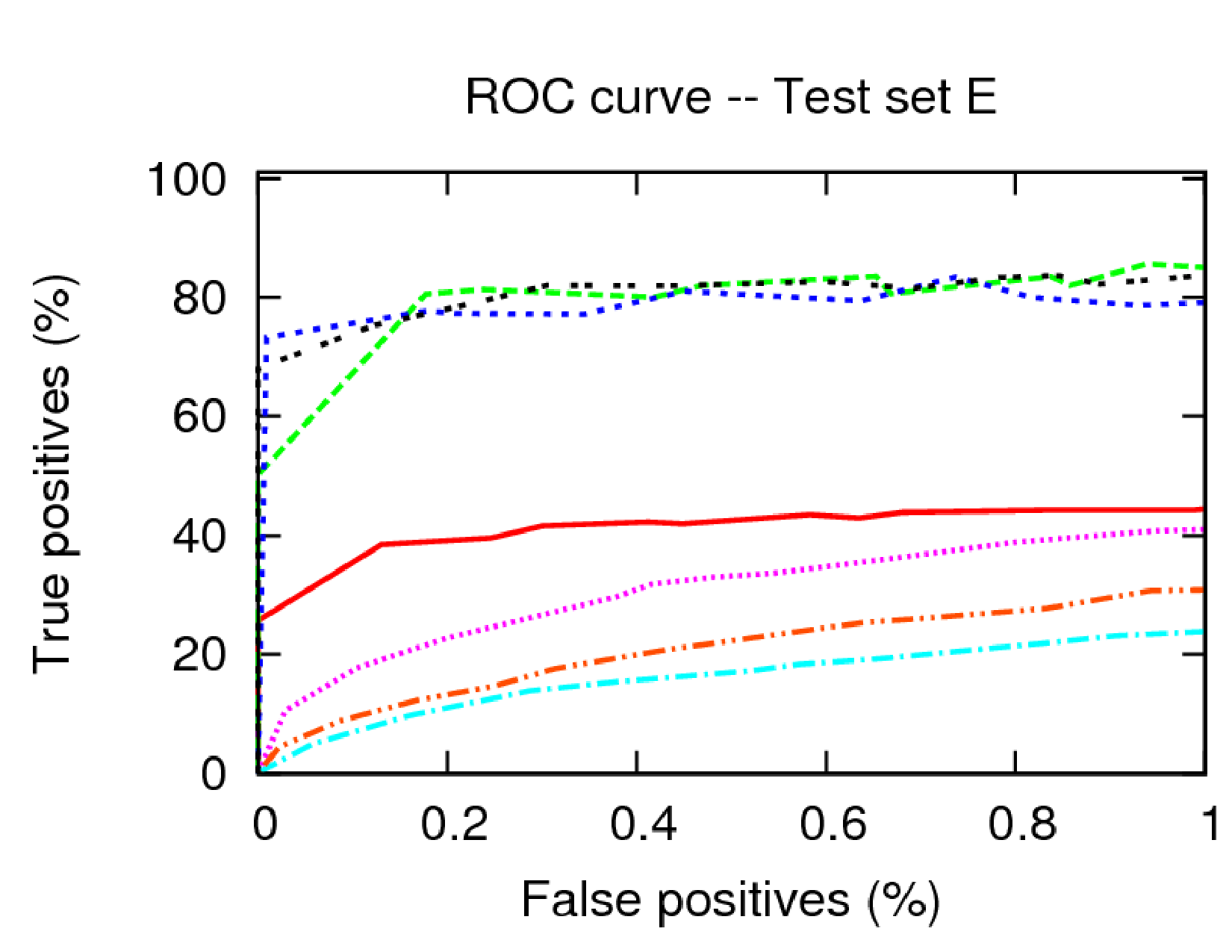}
  }
  \subfigure{
   \includegraphics[width=70mm]{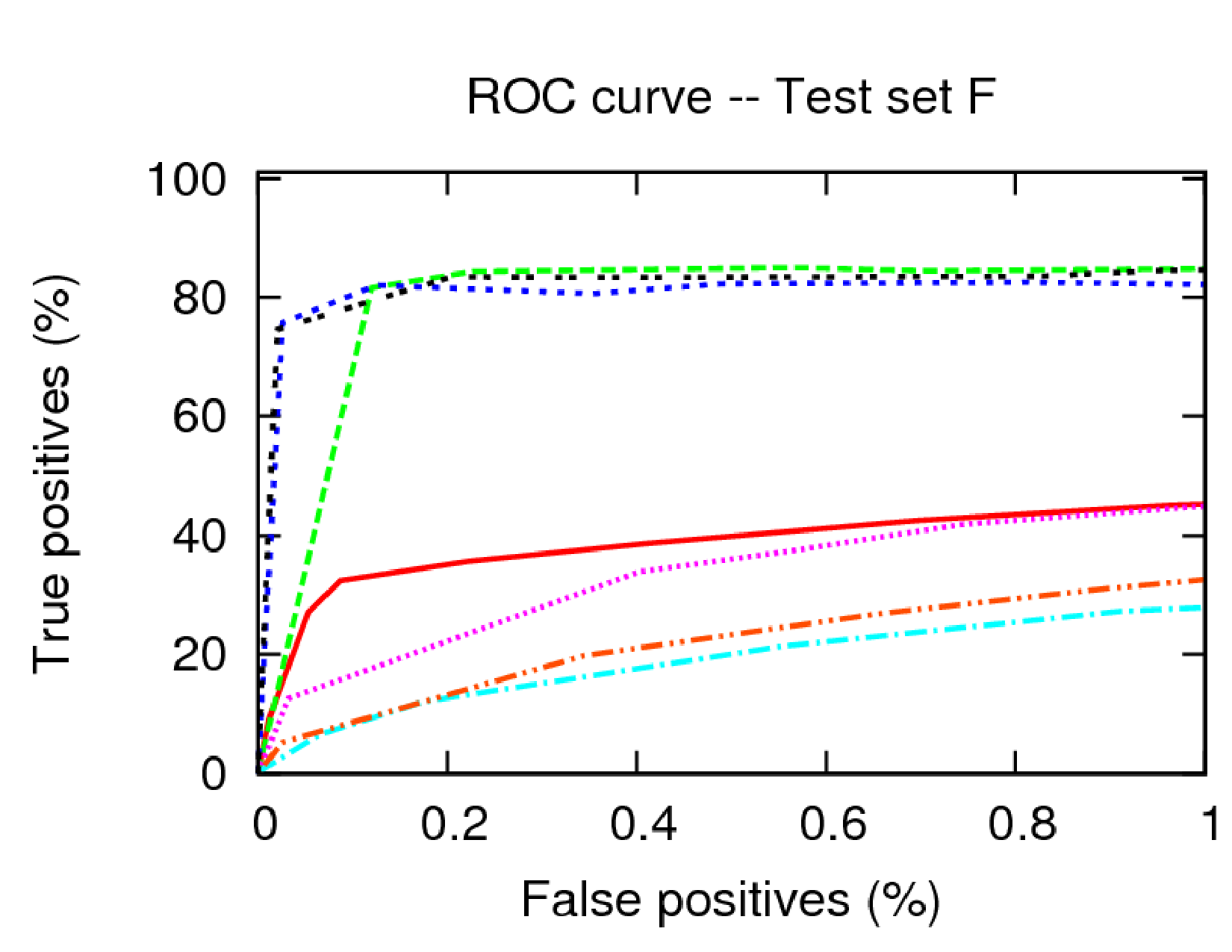}
  }
  \subfigure{
   \includegraphics[width=70mm]{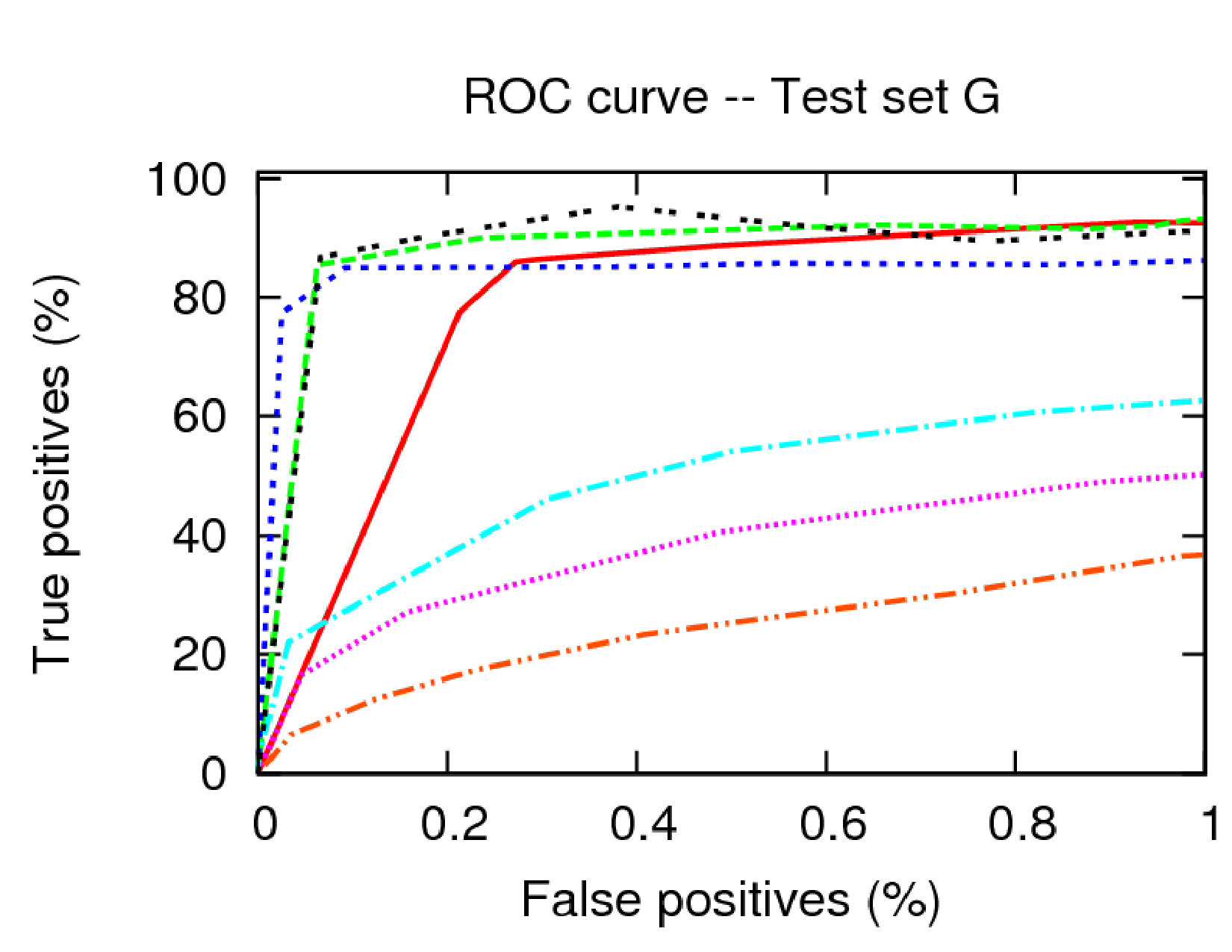}
  }
  \subfigure{
   \includegraphics[width=70mm]{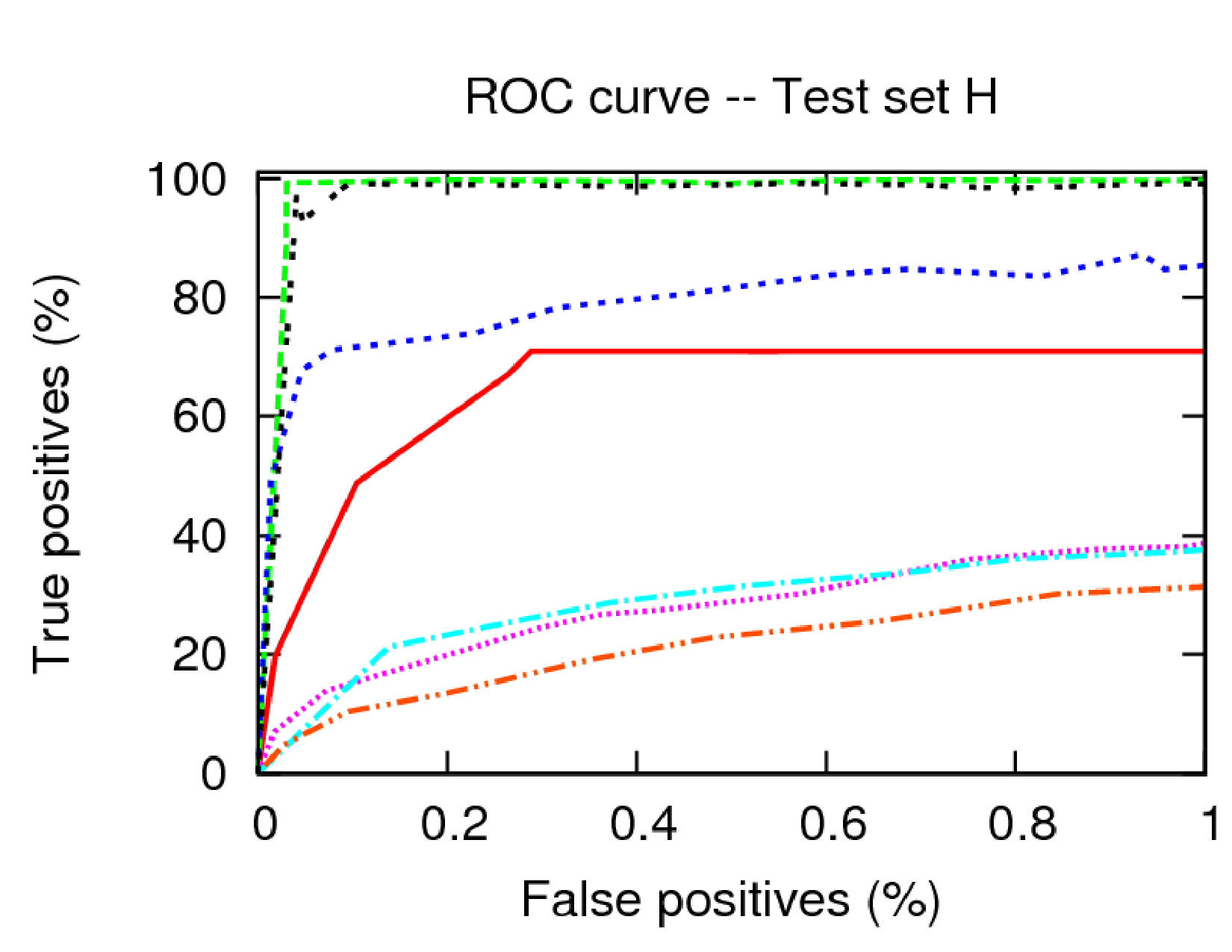}
  }
 \end{center}
 \caption{The ROC curves produced by applying various RFI detection methods to the test sets. The closer an ROC curve passes the top-left of the graph at 100\% true-positives with 0\% false-positives, the more accurate is the method. }
 \label{fig:roc-curves}
\end{figure*}

Both the SVD and threshold methods show accurate results on removing line RFI and broadband RFI. The SVD method is not suitable for removing frequency-varying RFI, as demonstrated in Figure~\ref{fig:svd-diagonal-line}, and thus has to be complemented with other techniques to remove all RFI. However, the SVD method can be used to subtract and remove the RFI from the image, leaving the astronomical signal intact. For this to be succesful, considerable assumptions about the mathematical properties of RFI and the astronomical signal have to be true: the time-frequency matrix with the RFI components has to be orthogonal to the time-frequency matrix of the astronomical signal, and the different RFI components have to be either orthogonal to each other or linearly dependent on each other. 
Figure~\ref{fig:testsetH-svd} shows the SVD decomposition of test set A that consists of uncorrelated noise and linear RFI.

As it is hard to quantitatively compare RFI mitigation methods based on data sets of which the characteristics of the RFI cannot be known for certain, several artificial test sets were created. These sets are shown in Figure~\ref{fig:testsets} and contain broadband RFI only. Since the RFI was added artificially, the location of the RFI in the time-frequency domain is known, and the accuracy of the methods can be tested quantitatively. The results are drawn as receiver operating characteristic (ROC) curves in Figure~\ref{fig:roc-curves}. ROC curves show the true probability rate against the false probability rate. The different accuracies and characteristics of the methods can easily be compared in ROC graphs.

The \texttt{SumThreshold} method shows a considerably better accuracy in all the test sets. Test sets A and B contain RFI that is completely linear dependent, and the SVD method also works very well in these sets. The SVD method could actually be used to subtract the RFI instead of flagging and not using the data. However, to mitigate the RFI in test set C, the methods have to deal with RFI that is neither orthogonal nor completely dependent on each other, and thus the accuracy of the SVD method decreases.

A normal thresholding strategy was also tested to compare the results. When performing normal thresholding with a surface fit as in the \texttt{SumThreshold} method, the accuracy for thresholding actually decreases in the test cases without an astronomical signal (see the curves labelled ``Fit + simple threshold'' in Figure~\ref{fig:roc-curves}). This is partially because the surface fit was optimised for the \texttt{SumThreshold} method. Furthermore, since the accuracy of the thresholding is not very good, the fit is influenced by the undetected RFI, causing more errors.

\begin{figure*}
 \begin{center}
  \subfigure[SVD performed on test set H (71.0\% recognized, 0.6\% false).]{
   \label{fig:testsetF-svd}
   \includegraphics[width=112mm]{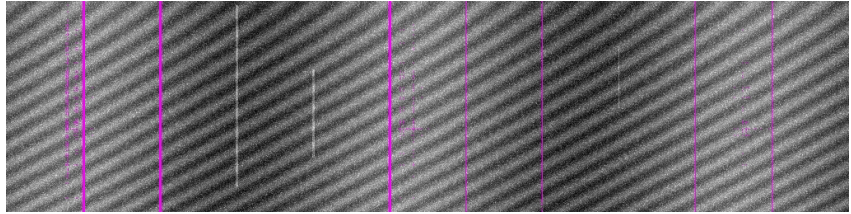}
   }
  \subfigure[\texttt{SumThreshold} performed on test set H (99.4\% recognized, 0\% false).]{
   \includegraphics[width=112mm]{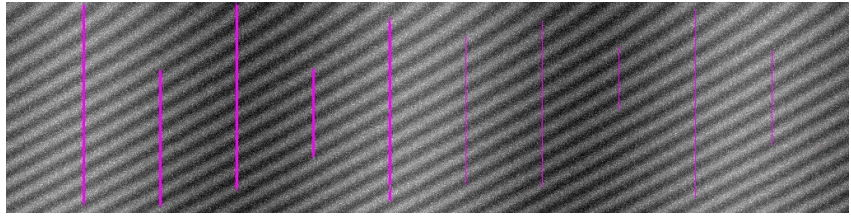}
   }
 \end{center}
 \caption{The results of two mitigation methods applied to test set H.}
 \label{fig:testsetF}
\end{figure*}

When astronomical information is added as in test set E and a more complex background is added as in test set F, the SVD method shows a decreased accuracy in mitigating the RFI, as can also be seen in Figure~\ref{fig:testsetF}. However, in test set G, the background of test set F is Gaussian smoothed and subtracted, as is done before thresholding. The SVD method now shows an improved accuracy, though still not as good as the \texttt{SumThreshold} method. Test set H shows that the linear dependency of the RFI is not the only requirement for succesful mitigation with the SVD method: the added RFI is completely linearly dependent in this test set, but the background is still causing low accuracies in the SVD method.

It should be noted that some of these test sets are measuring the theoretical accuracy of non-orthogonal, but not completely independent RFI contamination. As shown in \S\ref{svd-method}, this was the hardest case for the SVD method. When in practice the RFI does behave in an orthogonal or dependent manner, the results might be quite different. Nevertheless, it is unlikely that all RFI contaminations that are measured by different antennae at different times are always either linearly dependent or orthogonal. 

The presented test sets simulate a single baseline, whereas in a real measurement, the SVD method will exploit the correlation of RFI between different antennae. This will, however, also decrease the probability that all RFI is either orthogonal or linearly dependent.

\begin{figure*}
 \begin{center}
  \subfigure[Original]{
   \label{fig:wsrt-original}
   \includegraphics[width=68mm]{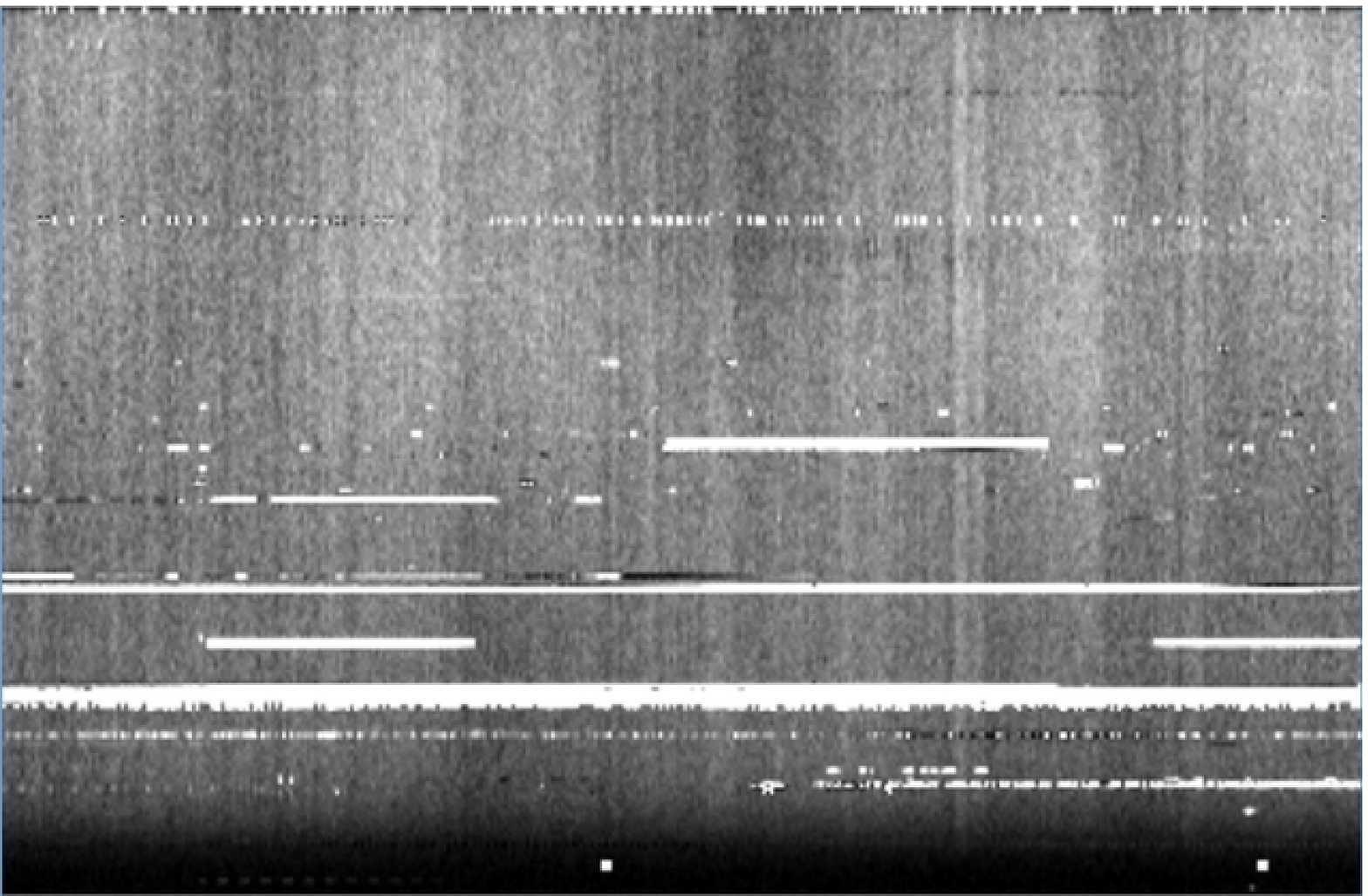}
   }
  \subfigure[Automated flagging result]{
   \label{fig:wsrt-thresholded}
   \includegraphics[width=68mm]{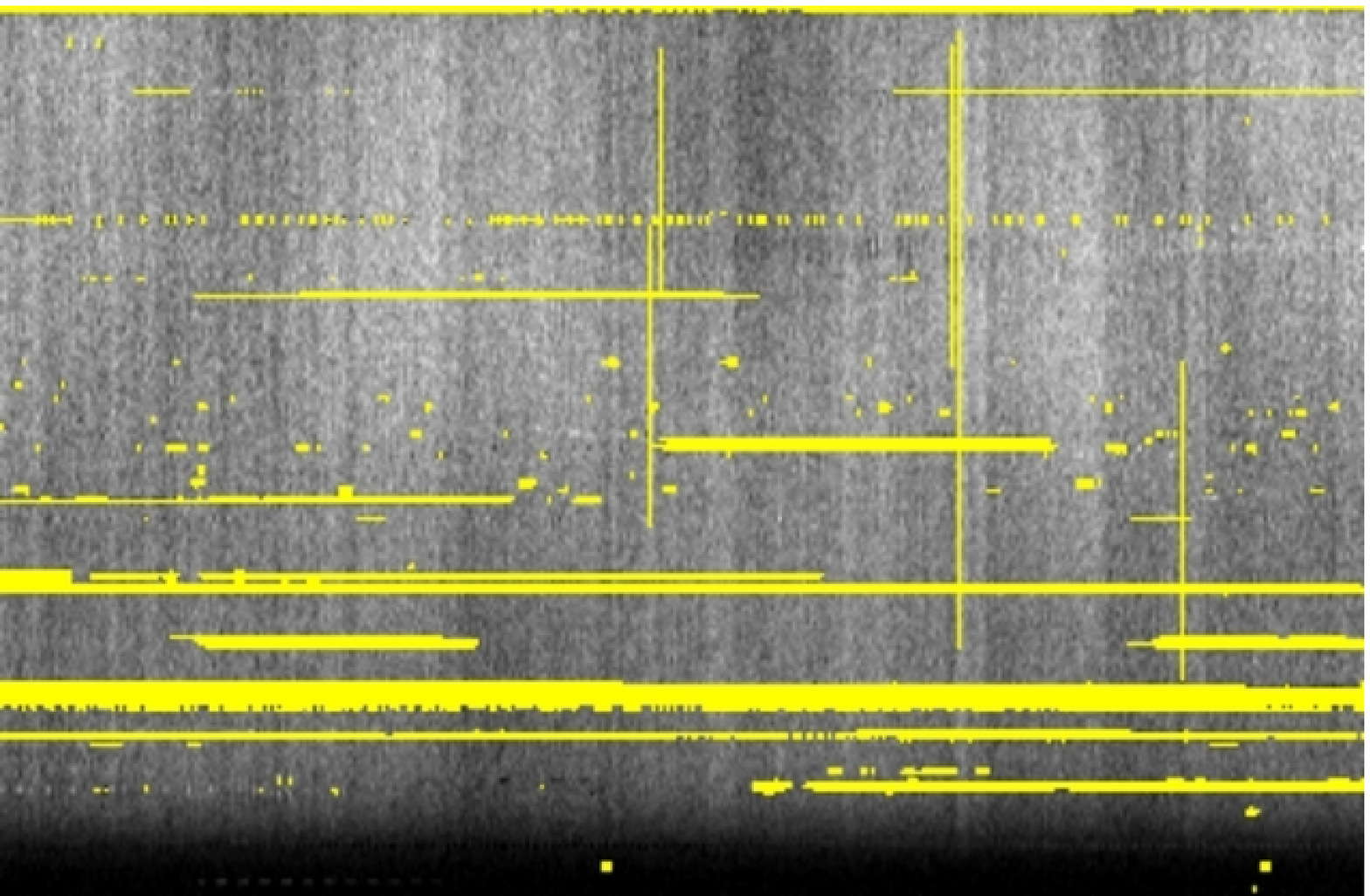}
   }
  \subfigure[Smoothed]{
   \label{fig:wsrt-smoothed}
   \includegraphics[width=68mm]{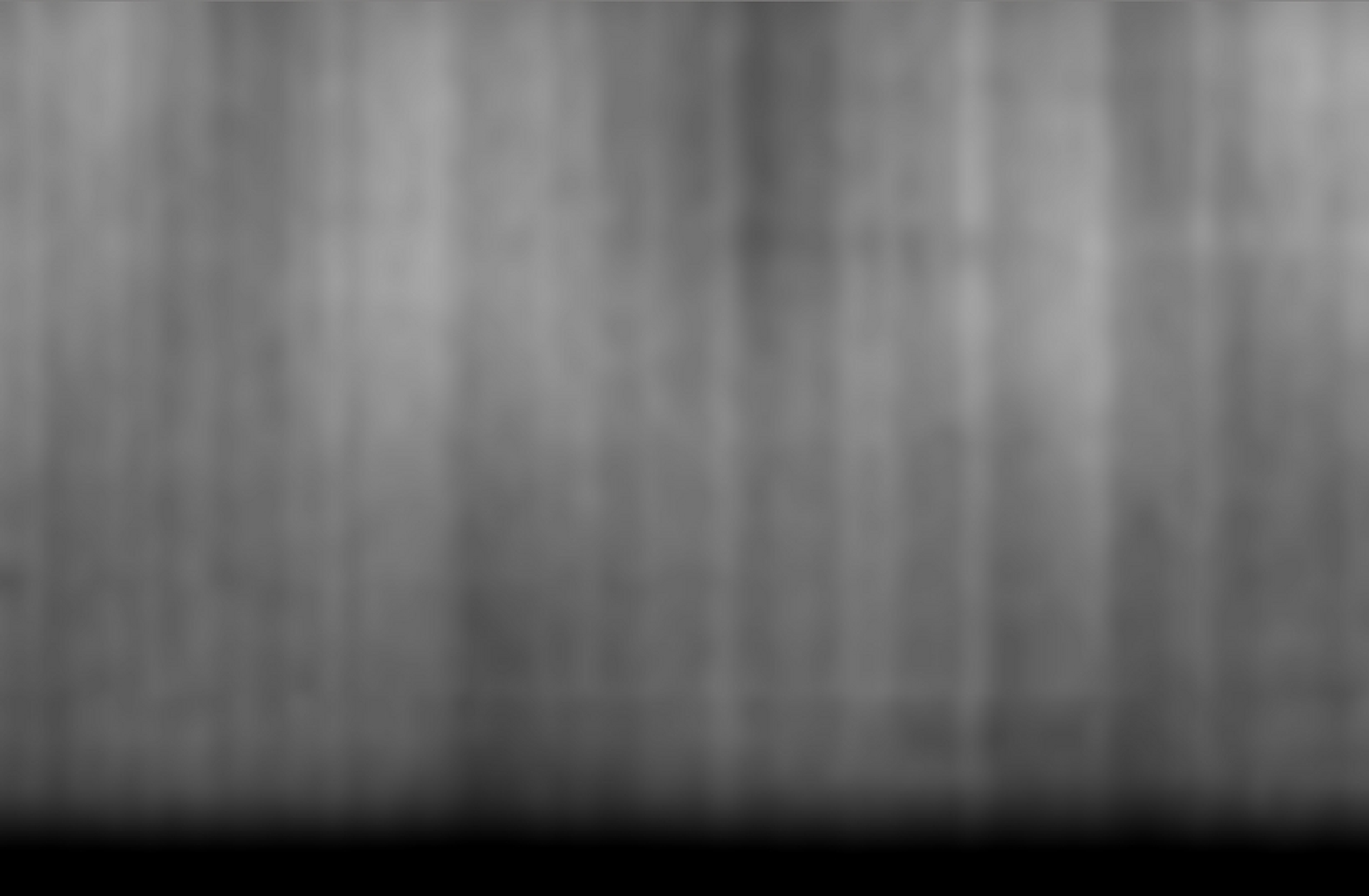}
   }
  \subfigure[Difference]{
   \label{fig:wsrt-difference}
   \includegraphics[width=68mm]{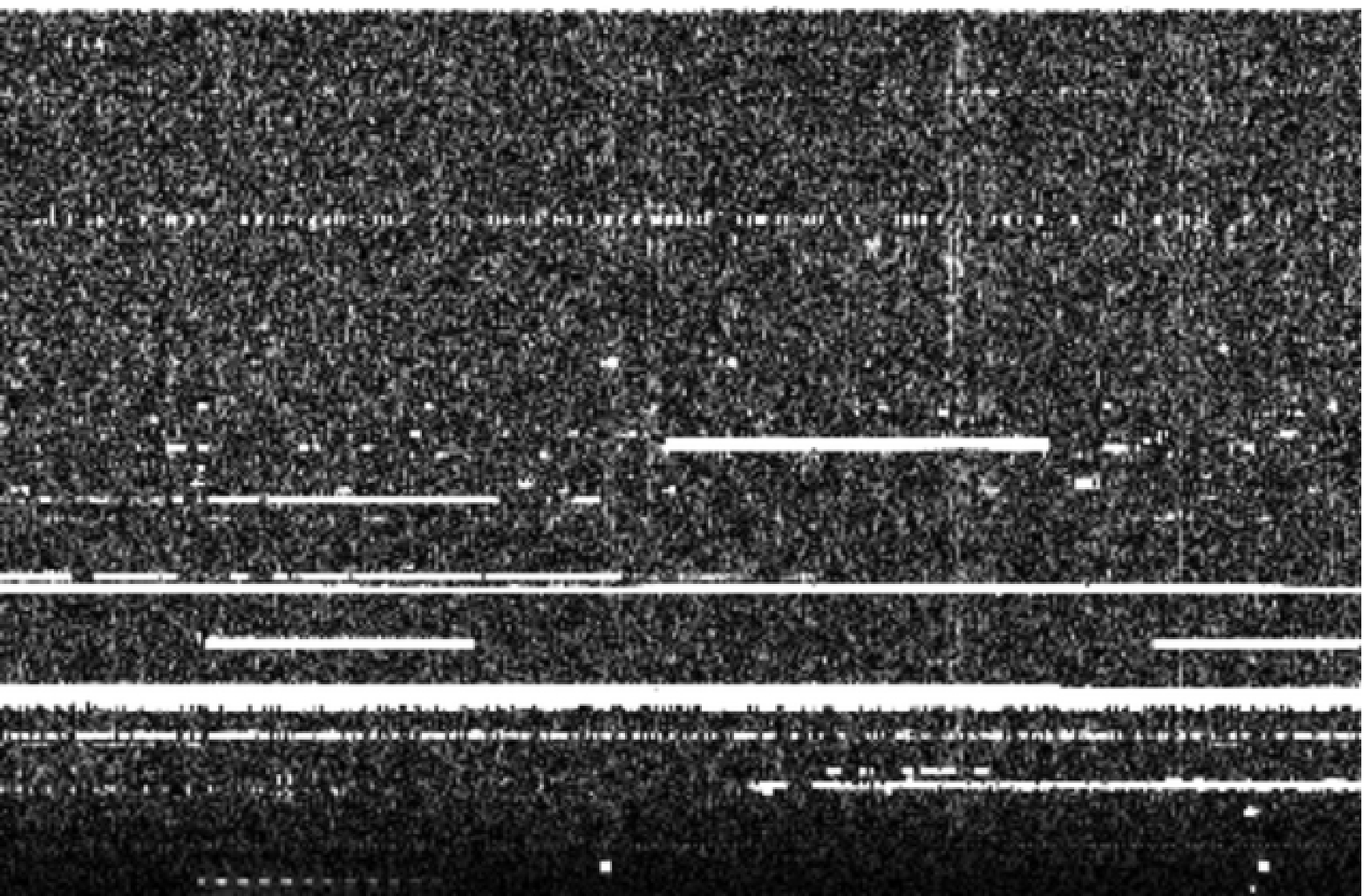}
   }
 \end{center}
 \caption{Time (horizontal) vs. frequency (vertical) plots of uncalibrated WSRT data, cross-correlations of antenna C vs. D. and the application of the \texttt{SumThreshold} automated flagging procedure. Panel \subref{fig:wsrt-original} shows one hour of the amplitude of a 3C196 observation, panel \subref{fig:wsrt-thresholded} shows the result of the flagger, panel \subref{fig:wsrt-smoothed} shows the fitted surface after 5 iterations, and panel \subref{fig:wsrt-difference} shows the difference between panel \subref{fig:wsrt-original} and panel \subref{fig:wsrt-smoothed}. }
 \label{fig:wsrt-crosscorrelations}
\end{figure*}

\subsection{Automatic flagging of WSRT data} \label{wsrt-results}

To test the various RFI flagging algorithms we have used WSRT data in
the LFFE band from 138-157 MHz obtained in November and December 2007.
The observations have been described and analysed by
\citet{bernardi-wsrt-foregrounds-1,bernardi-wsrt-foregrounds-2}
to which we refer for details of the astrophysical motivation and
calibration. For our analysis, however, we used the raw uncalibrated
visibilities. The correlator integration time for the data was
10s. A total of 8 bands of 2.5 MHz width were available. The central
frequencies of these bands were located at 139.3, 141.5, 143.7, 145.9,
148.1, 150.3, 152.5 and 154.7 MHz. Each band was divided into 512
spectral channels. The data were Hanning tapered, yielding an effective
spectral resolution of 9.8 kHz. Therefore, adjacent spectral channels are
highly correlated. A total of 13 telescopes participated in
the observations providing a total of 78 interferometers with
baselines from 36 to 2736 meters. All four cross-correlations
between the orthogonal, linearly polarized feeds were used in the
analysis.

We have tested the various methods on several data sets. The \texttt{SumThreshold} method in combination with Gaussian smoothing shows especially excellent results. Figure~\ref{fig:wsrt-crosscorrelations} shows a typical time-frequency diagram of WSRT data at $\sim$140 MHz and the application of the \texttt{SumThreshold} method. Although the smoothed surface is slightly affected by the RFI after five iterations, as faint artefacts are visible in the smoothed surface around places where RFI used to be, the effect is so small that it does not pose a problem for the \texttt{SumThresholding} method. However, it makes the calculated false probability rate inaccurate, as the false probability calculations assume independence between the residual samples. When validating the results by visual inspection, we see far less false detections than the calculated false probability rate.

\begin{figure}
 \begin{center}
  \subfigure[Original]{
   \label{fig:wsrt-3c147-original}
   \includegraphics[width=70mm]{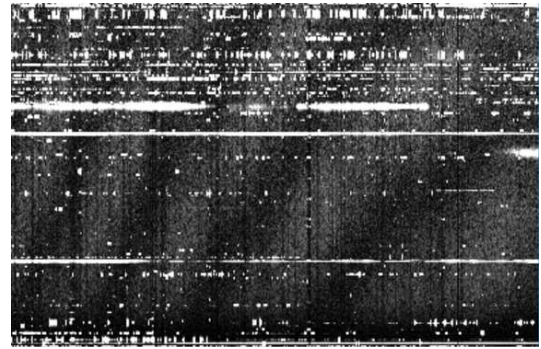}
   }
  \subfigure[Automated flagging result]{
   \label{fig:wsrt-3c147-smoothed}
   \includegraphics[width=70mm]{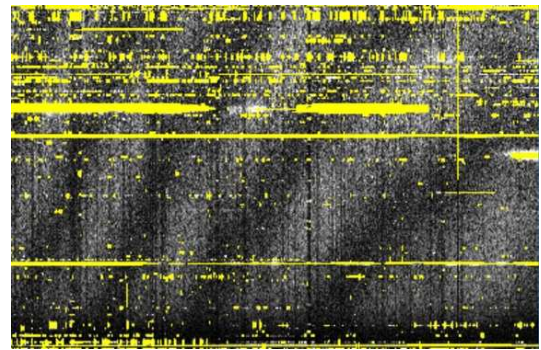}
   }
 \end{center}
 \caption{Time (horizontal) vs. frequency (vertical) plots of WSRT data, cross-correlations of antenna 1 vs. 2: a particular bad band at 121.3 MHz - 123.7 MHz of an observation of 3C147, showing that the method remains robust in one of the worst cases at the WSRT.}
 \label{fig:wsrt-3c147-crosscorrelations}
\end{figure}

We were able to use the same parameters for any situation in the WSRT data, and therefore were able to completely automate the flagging process. Even at baselines and frequencies with dramatic RFI contamination of up to 50\%, the \texttt{SumThreshold} flagging method remained stable and accurate. Figure~\ref{fig:wsrt-3c147-crosscorrelations} shows, for example, a badly contaminated band of WSRT data that is almost perfectly RFI flagged.

\section{Conclusion and discussion} \label{conclusion-chapter}
In this article we have shown several approaches to deal with RFI that is left after correlation. The results show that automated flagging with the \texttt{SumThreshold} method works well for broadband and peak RFI. In all cases, the default parameters for the method work well, although parameter tweaking might in some cases improve the classification. In the artificial broadband RFI situations, it detects 80\% of the artificially inserted RFI with less than 0.1\% error, and often approaches a 99\% recognition almost without error. The accuracy of this method is therefore as good as can be expected from manual flagging. In the case of WSRT, the new method does not improve the dynamic range of the data compared with manual flagging, but the method saves a considerable amount of work.

New telescopes such as LOFAR and SKA require robust automatic procedures, as these telescopes will produce data sets that exceed current measurements in volume by orders of magnitude. The ability to flag or check baselines or subbands individually will be lost.

The ROC analysis shows that the \texttt{SumThreshold} method is to be preferred above the \texttt{VarThreshold} and SVD methods. The SVD method can be used in some respects to detect RFI, but is less accurate. It can either be used to detect the RFI or to correct samples. If it is used to correct samples by filtering the RFI out, rather
than only detecting and flagging it, artefacts with unknown characteristics could remain in the data. For WSRT data, these artefacts look as bad as the broadband RFI itself.

All methods have been tested without assuming a data model. Subtracting the model before RFI detection might improve the classification further. Nevertheless, the detection accuracy with and without a model do not differ much. As such, going back and forth between flagging data and creating a model is not necessary in most cases.

\section{Further work} \label{future-work-chapter}
RFI with a moderate strength that can be detected by eye was found to be of no concern for automatic flagging methods in sensitive telescopes such as WSRT. However, a different kind of RFI might still pose problems. Certain weak RFI, such as radiation that leaks from cabins in situ, might be present in many channels for a substantial duration of the observation. This might pose problems for observations that require long integration times to achieve their required signal-to-noise ratios, such as the LOFAR-EoR project. If the RFI is persistent in time, systematic errors could result. There are some interesting ways to remove these, and one of them is the fringe-fitting RFI mitigation method described by \citet{fringe-fitting-rfi-mitigation}. Although this technique works at the GMRT, preliminary tests with the fringe-fitting RFI mitigation method on WSRT and LOFAR data do not show a strong presence of this type of RFI, and removing very weak RFI with a similar method requires more work. Therefore, to determine whether this type of RFI is really present, and whether it might be removable is yet to be seen.

An important next step is to consider practical issues in RFI mitigation techniques. For example, the effects of many RFI mitigation methods, post as well as pre-correlation, need to be simulated, since we never know what the image plane ought to look like. Also, which post and pre-correlation methods can be combined? Under which practical circumstances do RFI mitigation methods fail? How can we be sure that astronomical detections are not caused by RFI, or by the methods that try to reduce RFI? Answering these questions is important for establishing the reliability of new RFI mitigation methods and for their regular use by astronomers. 

\begin{figure}
 \begin{center}
  \includegraphics[width=80mm]{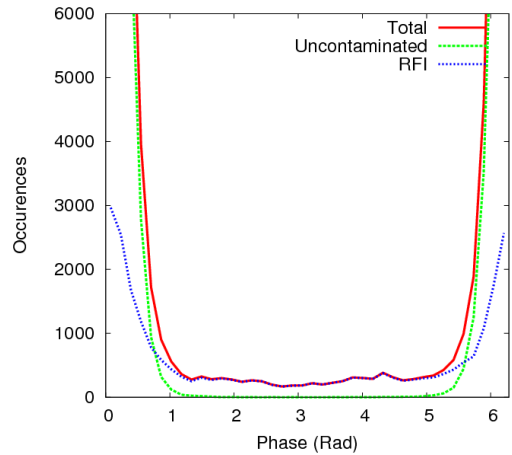}
 \end{center}
 \caption{Typical histogram of the phase in a short baseline of a WSRT observation. The RFI was detected by using the \texttt{SumThreshold} method. The plot implies that RFI-contaminated samples have a much higher probability to have a phase deviating from zero, and the phase thus contains valuable information for RFI detection. }
 \label{fig:phase-distribution}
\end{figure}

Although, at this point, it seems to be of little concern to improve the \texttt{SumThreshold} automatic flagging method any further, it might be interesting to improve it by combining more information for detection and by using fuzzy logic to decide the sample classification. An interesting example would be to include phase information in the recognition, as only the amplitude information has been used so far by the threshold methods. For example, Figure~\ref{fig:phase-distribution} shows that the phase contains valuable information about a sample: in uncontaminated samples, the phase is likely to be near zero rotation, whereas many contaminated samples do have a phase deviating from zero. Other distinguishing information could be contained in the polarization information per sample and in the combination of different baselines.

Based on the low frequency observations with the WSRT, it can be expected that the radio environment of LOFAR is sufficiently clean for sensitive astronomical experiments. In a future paper we will fully analyse and describe the LOFAR environment and the effectiveness of the RFI strategies.

Finally, we would like to emphasise that the methodology of RFI flagging, or any kind of error detection, needs to change because of the introduction of telescopes such as LOFAR, that generate so much data that it is not possible for astronomers to browse through the data for ``the baseline that was producing this artefact'' or ``the timestep that corresponds to these stripes in my image''. Therefore, another important next step is to be able to automatically detect errors that are caused by RFI, calibration issues, broken hardware, faulty software or any step in the complicated pipeline of a radio observatory.

\section*{Software}
Software to flag measurement sets with the \texttt{SumThreshold} method and other discussed methods has been made publicly available and can be downloaded from the following location:\\
\url{http://www.astro.rug.nl/rfi-software/}
\section*{Acknowledgement} 
A.G. de Bruyn thanks Peter Fridman for many discussions on RFI mitigation.

\bibliographystyle{mn2e}
\bibliography{RFI-Classification-Offringa}

\label{lastpage}

\end{document}